\documentclass{jnmp}

\usepackage{amsmath}

\JNMPnumberwithin{equation}{section}

\begin{document}

\renewcommand{\evenhead}{U Mu\u{g}an and F Jrad}
\renewcommand{\oddhead}{Painlev\'{e} Test and Higher Order Differential Equations}

\thispagestyle{empty}

\FirstPageHead{9}{3}{2002}{\pageref{mugan-firstpage}--\pageref{mugan-lastpage}}{Article}

\copyrightnote{2002}{U Mu\u{g}an and F Jrad}

\Name{Painlev\'{e} Test and Higher Order\\ Differential Equations}
\label{mugan-firstpage}

\Author{U\u{g}urhan MU\u{G}AN~$^{\dag\star}$
and Fahd JRAD~$^\ddag$}

\Address{$^\dag$~SUNY at Buffalo, Department of Mathematics, 305 Mathematics Bldg.,\\
~~Buffalo, NY 14260-2900 U.S.A. \\
~~E-mail: umugan@acsu.buffalo.edu, \\[10pt]
$^\star$~Permanent Address: Bilkent University, Department
of Mathematics,\\
~~06533 Bilkent, Ankara, Turkey\\
~~E-mail: mugan@fen.bilkent.edu.tr\\[10pt]
$^\ddag$ Cankaya University, Department of Mathematics  and Computer Sciences,\\
~~06530 Cankaya, Ankara, Turkey \\
~~E-mail: fahd@cankaya.edu.tr}

\Date{Received September 16, 2001; Revised November 11, 2001;
Accepted April 15, 2002}

\begin{abstract}
\noindent
Starting from the second Painlev\'{e} equation, we obtain
Painlev\'{e} type equations of higher order by using the singular point analysis.
\end{abstract}

\section{Introduction}

Painlev\'{e} and his school \cite{Painleve,Painleve1,gambier,ince}
investigated  second-order first-degree equations of the form
\begin{equation}
  {y''}=F(z,y,{y'}),
\label{eq:1}
\end{equation}
where $F$ is rational in $y'$, algebraic in $y$ and locally analytic in
$z$, and has no movable critical points.
This property is known as the Painlev\'{e}
property and ordinary differential equations (ODE), which possess it, are said
to be of Painlev\'{e} type.  They found that, within a M\"{o}bius
transformation, there exist fifty such equations.
Distinguished among these fifty equations are the six
Painlev\'{e} equations, PI, \dots, PVI. Any other of the fifty equations either
can be integrated in terms of known functions or can be reduced
to one of these six equations.

Higher order first-degree in the polynomial class and higher order higher degree
equations of Painlev\'{e} type  were investigated by Fuchs~\cite{fuchs, ince},
Briot and Bouquet~\cite{ince}, Cha\-zy~\cite{cazy},
Bureau~\cite{bur3,bur3a}, Exton~\cite{exton},
Martynov~\cite{marty}, Cosgrove~\cite{cos1, cos2}, Kudryashov~\cite{kudr}, Clarkson,
Joshi and Pickering~\cite{clark} and
also in the articles~\cite{mug-sakka, mug-sakka1, mug-sakka2, mugan-fahd}.
The Riccati equation is the only example for the first-order first-degree
equation which has the Painlev\'{e} property.
The best known third order
equation is Chazy's natural-barrier equation
\begin{equation}
y'''=2yy''-3{y'}^{2}+\frac{4}{36-n^2}\left(6y'-y^2\right)^{2},
\label{eq:2}
\end{equation}
The case $n=\infty$ appears in several physical problems. The equation
(\ref{eq:2}) is integrable for all real and complex $n$ and $n=\infty$.
Its solutions are rational for real $n$ in $2 \leq n \leq 5$, and have a circular
natural barrier for $n \geq 7$ and $n=\infty$.

In this article the second Painlev\'{e} hierarchy is investigated by
using the Painlev\'{e} ODE test or  singular point analysis.
It is possible to obtain the
equation of Painlev\'{e} type of any order, as well as
the known ones, starting from the second Painlev\'{e} equation.
Painlev\'{e} ODE test which is an algorithm
introduced by Ablowitz, Ramani, Segur (ARS)~\cite{ablow, ablow1}
tests whether a given ODE satisfies
the necessary conditions to be of Painlev\'{e} type.

The procedure to obtain higher order Painlev\'{e} type equations
starting from the second Painlev\'{e} equation may be summarized as follows:

{\bf I.} Take an $n^{th}$ order Painlev\'{e} type differential equation
\begin{equation}
y^{(n)}=F\left(z,y,y',\ldots,y^{(n-1)}\right),
\label{eq:3}
\end{equation}
where $F$ is analytic in $z$ and rational in its other arguments.
If $y \sim y_{0}(z-z_{0})^{\alpha}$ as $z \rightarrow z_0$,
then $\alpha $ is a negative integer for certain values of $y_{0}$.
Moreover, the highest derivative
term is one of the dominant terms. Then the dominant terms are of order
$\alpha -n$. There are $n$ resonances $r_{0}=-1, r_{1},r_{2},\ldots,r_{n-1}$,
with all $r_{i}$, $i=1,2,\ldots,(n-1)$ being nonnegative real distinct integers
such that $Q(r_{j})=0$, $j=0,1,2,\ldots,(n-1)$. The compatibility conditions
for the simplified equation that retains only dominant terms of~(\ref{eq:3})
are identically satisfied. Differentiation of the simplified equation
with respect to~$z$ yields
\begin{equation}
y^{(n+1)}=G\left(z,y,y',\ldots,y^{(n)}\right),
\label{eq:4}
\end{equation}
where $G$ contains the terms  of order $\alpha-n-1$, and the
resonances of (\ref{eq:4}) are the roots of $Q(r_{j})(\alpha +r-n)=0$.
Hence equation (\ref{eq:4}) has a resonance
$r_{n}=n-\alpha$ additional to the resonances of (\ref{eq:3}).
Equation (\ref{eq:4}) passes the Painlev\'{e} test
provided that $r_{n} \neq r_{i}$,
$i=1,2,\ldots,(n-1)$ and is a positive integer. Moreover the compatibility
conditions are  identically satisfied, that is $z_{0}, y_{r_{1}},\ldots,
y_{r_{n}}$ are arbitrary.

{\bf II.} Add the dominant terms which are not contained in $G$. Then the
resonances of the new equation are the zeros of a polynomial $\tilde{Q}(r)$
of order $n+1$. Find the coefficients of~$\tilde{Q}(r)$ such that
there is at least one principal Painlev\'{e} branch, that is,
all $n+1$ resonances (except $r_0 =-1$) are real positive distinct
integers for at least one possible choice of $(\alpha, y_{0})$.
The other possible choices of $(\alpha, y_{0})$ may give the secondary
Painlev\'{e} branch, that is, all the resonances are distinct integers.

{\bf III.} Add the nondominant terms which are the terms of weight less than
$\alpha -n-1$, with  coefficients analytic in $z$.
Find the coefficients of the nondominant terms by the use of the compatibility
conditions and transformations which preserve the Painlev\'{e} property.

In this article we consider only the principal branch, that is,
all the resonances $r_{i}$ (except $ r_{0}=-1$) are positive real
distinct integers and the number of resonances is equal to the
order of the differential equation for a possible choice of
$(\alpha, y_{0})$. Then, the compatibility conditions give full
set of arbitrary integration constants. The other possible choices
of $(\alpha, y_{0})$ may give secondary branches which possess
several distinct negative integer resonances. Negative but
distinct integer resonances give no conditions which contradict
integrability~\cite{andrew2}. Higher order equations with negative
but distinct integer resonances might be investigated separately.
In the present work we start with the second Painlev\'{e} equation
and obtain the third, fourth and some fifth  order equations of
Painlev\'{e} type. All of the third order and some of the fourth
order equations are found  in the literature, but for the sake of
completeness we present the known equations with appropriate
references. A similar procedure was used in~\cite{mugan-fahd} to
obtain the higher order equations of  Painlev\'{e} type  by
starting from the first Painlev\'{e} equation. The procedure can
also be used to obtain the higher order equations by starting from
PIII, \dots, PVI. These results will be published elsewhere.

\section{Third order equations: $\boldsymbol{\mbox{P}_{\rm II}^{(3)}}$}

The second Painlev\'{e} equation, PII, is
\begin{equation}
y''=2y^{3}+zy+\nu.
\label{eq:26}
\end{equation}
The Painlev\'{e} test gives that there is only one branch and the resonances are $(-1,4)$.
The dominant terms of (\ref{eq:26}) are $y''$ and $2y^3$ which are of order $-3$ as
$z \rightarrow z_{0}$.
Differentiation of the simplified equation $y''=2y^3$ gives
\begin{equation}
y'''=ay^{2}y',
\label{eq:28}
\end{equation}
where $a$ is a constant which can be introduced by replacing $y$ with
$\lambda y$, such that $6\lambda^{2}=a$.
Addition of the polynomial type terms of order
$-4$  gives the following simplified equation
\begin{equation}
y'''=a_{1}yy''+a_{2}y'{}^{2}+a_{3}y^{2}y'+a_{4}y^{4},
\label{eq:29}
\end{equation}
where $a_{i}$, $i=1,\ldots,4$ are constants. Substitution of
\begin{equation}
y=y_{0}(z-z_{0})^{-1}+\beta (z-z_{0})^{r-1},
\label{eq:30}
\end{equation}
into the simplified equation (\ref{eq:29}) gives the following equations, $Q(r)=0$, for the
resonance~$r$ and
for $y_{0}$, respectively,
\begin{subequations}
\begin{gather}
{Q}(r)=(r+1)\left\{r^{2}-(a_1 y_0 +7)r-\left[a_3 y_{0}^2-2(2a_1 +a_2)y_0 -18\right]\right\}=0,
\label{eq:31a}\\
a_4 y_{0}^3-a_3 y_{0}^2+(2a_1+a_2)y_0 +6=0.
\label{eq:31b}
\end{gather}\label{eq:31}
\end{subequations}
Equation (\ref{eq:31}b) implies that, in general, there are three
branches of Painlev\'{e} expansions if $a_{4}\neq 0$. Now one
should determine $y_{0j}$, $j=1,2,3$, and $a_i$ such that at least
one of the branches is the principal branch. There are three cases
which should be considered separately.

{\bf Case I.} $a_{3}=a_{4}=0$: In this case $y_0$ takes a single value, i.e.\
 there is only one branch.
 Equation (\ref{eq:31}a) implies that
$r_{0}=-1$ and $r_{1}r_{2}=6$. Therefore there are the following four possible cases:
\begin{gather}
\mbox{{\bf a.}} \ \ y_{01}=-6/a_2~:~(r_{1},r_{2})=(1,6),\quad a_{1}=0,\nonumber\\
\mbox{{\bf b.}} \ \ y_{01}=-2/a_1~:~(r_{1},r_{2})=(2,3),\quad a_{1}=a_{2},\nonumber\\
\mbox{{\bf c.}} \ \ y_{01}=-12/a_1:~(r_{1},r_{2})=(-2,-3), \quad a_{1}=-2a_{2}/3,\nonumber\\
\mbox{{\bf d.}} \ \ y_{01}=-14/a_1:~(r_{1},r_{2})=(-1,-6).
\label{eq:33}
\end{gather}
The case d is not be considered since $r=-1$ is a double resonance.
The compatibility conditions are identically satisfied for the first two cases.
To find the canonical form of the third-order equations of Painlev\'{e} type, one
should add nondominant terms  the coefficients of which are analytic functions of $z$,
that is, one should consider the following equation for each case
\begin{equation}
y'''=a_{1}yy''+a_{2}y'{}^{2}+A_{1}y''+A_{2}yy'+A_{3}y^{3}+A_{4}y'+A_{5}y^{2}+A_{6}y+A_{7},
\label{eq:34}
\end{equation}
where $A_{k}(z)$, $k=1,\ldots,7$, are analytic functions of $z$.
Substitution of
\begin{equation}
y=y_{0}(z-z_{0})^{-1}+\sum_{j=1}^{6}y_{j}(z-z_{0})^{j-1},
\label{eq:35}
\end{equation}
into equation (\ref{eq:34}) gives the recursion relation  for $y_j$. Then
one can find $A_k$ such that the recursion relation, i.e.\ the compatibility
conditions for $j=r_1 , r_2$,
are identically satisfied and hence $y_{r_1}$, $y_{r_2}$ are arbitrary.

{\bf I.a:} The Painlev\'{e} property is preserved under the following
linear transformation
\begin{equation}
y(z)=\mu (z) u(t)+ \nu(z), \qquad t=\rho (z),
\label{eq:36}
\end{equation}
where $\mu$, $\nu$ and $\rho$ are analytic functions of $z$.
By using the transformation
(\ref{eq:36}) one can set
$A_4=A_5$, $A_1=0$, and $a_{2}=-6$.
The compatibility condition at the resonance $r_{1}=1$ gives $A_{2}=A_3$. The arbitrariness
of $y_{1}$ in the recursion relation for $j=6$ and the recursion relation yield that
\begin{equation}
A_{5}''-A_{5}^{2}=0, \qquad A_{6}''-A_{5}A_{6}=0,\qquad
A_{7}''-\frac{1}{3}A_{5}A_{7}=\frac{1}{6}A_{6}^{2},\qquad A_{3}=0.
\label{eq:38}
\end{equation}
According to the equation (\ref{eq:38}a) there are three cases should be considered
separately.

{\bf I.a.i:} $A_{5}=0$: From equation (\ref{eq:38}) all coefficients $A_{k}$ can be determined
uniquely.
The canonical form of the third order equation for this case is
\begin{equation}
y'''=-6y'{}^{2}+(c_1 z+c_2 )y+\frac{1}{72}c^{2}_{1}z^{4}+\frac{1}{18}c_{1}c_{2}z^{3}+
\frac{1}{12}c^{2}_{2}z^{2}+c_{3}z+c_{4},
\label{eq:40}
\end{equation}
where $c_{i}$, $i=1,\ldots,4$ are constants.

If $c_1 =c_2 =0$, then  (\ref{eq:40}) can be written as
\begin{equation}
u''=6u^{2}-c_{3}z-c_{4},
\label{eq:40a}
\end{equation}
where $u=-y'$. If $c_{3}=0$, then the solution of (\ref{eq:40a}) can be written in terms of
elliptic functions. If $c_{3} \neq 0$, (\ref{eq:40a}) can be transformed into the
first Painlev\'{e} equation.

If $c_1 =0$ and $c_2 \neq0$, (\ref{eq:40}) takes the following form by replacing $y$
by $\gamma y$ and $z$ by $\delta z$
such that $\gamma \delta =1$ and $c_2 \delta ^3=6$
\begin{equation}
y'''=-6y'{}^{2}+6y+3z^{2}+\tilde{c}_{3}z+\tilde{c}_{4},
\label{eq:41}
\end{equation}
where $\tilde{c}_{3}=c_{3}\delta ^{5}$, $\tilde{c}_{4}=c_{4}\delta ^{4}$.
Equation (\ref{eq:41}) was  given  in~\cite{cazy} and~\cite{bur3}.

If $c_1 \neq 0$ and $c_2 =0$, replacement of $y$
by $\gamma y$ and $z$ by $\delta z$ in (\ref{eq:40})
such that $\gamma \delta =1$ and $ c_1 \delta ^4=12$ yields
\begin{equation}
y'''=-6y'{}^{2}+12zy+2z^{4}+\tilde{c}_{3}z+\tilde{c}_{4},
\label{eq:42}
\end{equation}
where $\tilde{c}_{3}=c_{3}\delta ^{5}$, $\tilde{c}_{4}=c_{4}\delta ^{4}$.
Equation (\ref{eq:42}) was  given by Chazy~\cite{cazy} and Bureau~\cite{bur3}.
It should be noted that (\ref{eq:40}) can be reduced to (\ref{eq:42})
by the replacement of $z$ by $z-(c_{2}/c_{1})$
and then replacing  $y$ by $\gamma y$ and $z$ by $\delta z$
such that $\gamma \delta =1$, $c_1 \delta ^4=12$.

{\bf I.a.ii:} $A_{5}=\frac{6}{(z+c)^2}$: Without loss of generality the integration
constant $c$ can be set to zero.

From (\ref{eq:38}) the coefficients $A_{k}$ can be determined and
the canonical form of the equation is
\begin{gather}
 y'''=-6y'{}^{2}+6z^{-2}\left(y'+y^2\right)+\left(c_{1}z^{3}+c_{2}z^{-2}\right)y+
c_{3}z^2+c_{4}z^{-1} \nonumber\\
\phantom{y'''=} {}+\frac{1}{18}\left(\frac{1}{18}c^{2}_{1}z^8
+\frac{3}{2}c_{1}c_2 z^{3}+\frac{3}{4}c^{2}_{2}z^{-2}\right),
\label{eq:45}
\end{gather}
where $c_{i}$, $i=1,\ldots,4$, are constants.

If $c_1 =c_2 =0$, (\ref{eq:45})
is a special case of the equation given by Chazy~\cite{cazy}.
If $c_1 =0$ and $c_2 \neq0$, (\ref{eq:45}) takes the following form by replacing $y$
by $\gamma y$ and $z$ by $\delta z$ such that $\gamma \delta =1$ and $c_2 \delta=24$
\begin{equation}
y'''=-6y'{}^{2}+6z^{-2}\left(y'+y^2 +4y\right)+\tilde{c}_3z^2+
\tilde{c}_4z^{-1}+24z^{-2},
\label{eq:46}
\end{equation}
where  $\tilde{c}_{3}=c_{3}\delta ^{6}$
and $\tilde{c}_{4}=c_{4}\delta ^{3}$. Equation (\ref{eq:46}) is given in~\cite{bur3}.

If $c_1 \neq 0$ and $c_2 =0$, then equation (\ref{eq:45}) takes the form
\begin{equation}
y'''=-6y'{}^{2}+\frac{6}{z^2}(y'+y^2)+18z^3 y+z^8+
\tilde{c}_3z^2+\tilde{c}_4\frac{1}{z},
\label{eq:47}
\end{equation}
where $\tilde{c}_{3}$, $\tilde{c}_{4}$ are constants.
Equation (\ref{eq:47}) was  given in~\cite{bur3}.

{\bf I.a.iii:} If one replaces  $A_5$ with $6 \hat {A}_5$,  $A_6$ with $6 \hat {A}_6$ and
$A_7$ with $6 \hat {A}_7$, then  equations~(\ref{eq:38}) yields
\begin{equation}
\hat {A}_{5}''-6\hat{A}^{2}_{5}=0,\qquad \hat{A}_{6}''-6\hat{A}_{5}\hat{A}_{6}=0,\qquad
\hat{A}_{7}''-2\hat{A}_{5}\hat{A}_{7}=\hat{A}_{6}^{2}.
\label{eq:48}
\end{equation}
Integration of  (\ref{eq:48}a) once gives
\begin{equation}
 (\hat{A}'_{5})^{2}=4\hat{A}^{3}_{5}-\alpha_1,
\label{eq:49}
\end{equation}
where $\alpha_1$ is an integration constant. Then
\begin{equation}
\hat {A}_{5}={\mathcal P}(z;0,\alpha_1),
\label{eq:50}
\end{equation}
where ${\mathcal P}$ is the Weierstrass elliptic function.
If $\hat{A}_6=0$,  (\ref{eq:48}c)
implies that $\hat{A}_7$ satisfies  Lam\'{e}'s equation. Hence
\begin{equation}
\hat{A}_{7}=c_1 E_{1}(z)+c_{2}F_{1}(z),
\label{eq:50a}
\end{equation}
where $c_1$ and $c_2$ are constants,
$E_{1}(z)$ and $F_{1}(z)$ are the Lam\'{e} functions of degree one
of the first and second kind, respectively. They are given as
\begin{equation}
E_{1}(z)=e^{-z\zeta (a)}\frac{\sigma(z+a)}{\sigma(z)},\qquad
           F_{1}(z)= e^{z\zeta (a)}\frac{\sigma(z-a)}{\sigma(z)},
\label{eq:51}
\end{equation}
where $\zeta$ is the $\zeta$-Weierstrass function such that
$\zeta '=-{\mathcal P}(z)$, $\sigma$
is the  $\sigma$-Weierstrass function such that $\frac{\sigma'(z)}{\sigma (z)}=\zeta (z)$
and $a$ is a parameter such that ${\mathcal P}(a;0,\alpha_1)=0$.  Then the equation
\begin{equation}
y'''=-6y'{}^{2}+6{\mathcal P}(z;0,\alpha_1)\left(y'+y^{2}\right)+\tilde{c}_1 e^{-z\zeta (a)}
\frac{\sigma(z+a)}{\sigma(z)}
           +\tilde{c}_{2} e^{z\zeta (a)}\frac{\sigma(z-a)}{\sigma(z)},
\label{eq:52}
\end{equation}
where $\tilde{c}_1=6c_1$ and $\tilde{c}_2=6c_2$. Equation
(\ref{eq:52}) was  considered in~\cite{cazy}. If $\hat{A}_6 \neq
0$, ${\mathcal P}(z;0,\alpha_1)$ also solves  equation
(\ref{eq:48}b). Then (\ref{eq:48}c) implies that $\hat{A}_7$
satisfies the nonhomogeneous Lam\'{e}'s equation. Hence,
\begin{equation}
\hat{A}_7(z)=k_{1}(z)E_{1}(z)+k_{2}(z)F_{1}(z),
\label{eq:52a}
\end{equation}
where
\begin{equation}
k_{1}(z)=k_{1}-\int_{z}\frac{{\mathcal P}^{2}(t;0,\alpha_{1})}{W(t)}F_{1}(t) dt,\qquad
k_{2}(z)=k_{2}+\int_{z}\frac{{\mathcal P}^{2}(t;0,\alpha_{1})}{W(t)}E_{1}(t) dt
\label{eq:52b}
\end{equation}
with $k_{1}$ and $k_{2}$ are  constants of integration
and $W(z)=E_{1}F'_{1}-E'_{1}F_{1}$.

{\bf I.b:}
 The coefficients $A_k (z)$, $k=1,\ldots,7$, of the nondominant terms can be
found by using the  linear transformation (\ref{eq:36}) and the compatibility conditions.
The linear transformation (\ref{eq:36}) allows one to set
$a_{2}=-2$, $A_1 (z)=0$, $A_2 (z)=A_3 (z)$ and the compatibility conditions give that
$A_2 (z)=A_{6}(z)=0$ and $A_4 (z)=A_5 (z)$.
So the canonical form of the equation is
\begin{equation}
y'''=-2\left(yy''+y'{}^{2}\right)+A_4 \left(y'+y^2\right)+A_7 ,
\label{eq:56}
\end{equation}
where $A_4 $ and $A_7$ are arbitrary analytic functions of $z$. If one lets
$u=y'+y^{2}$,
then (\ref{eq:56}) can be reduced to a linear equation for $u$.
Equation (\ref{eq:56}) was  given in~\cite{bur3}.

{\bf I.c:} Without loss of generality one can choose $a_1 =2$. Then the simplified equation is
\begin{equation}
y'''=2yy''-3y'{}^{2},
\label{eq:58a}
\end{equation}
which was also considered in~\cite{cazy, bur3}.

{\bf Case II.} $a_4 =0$: In this case $y_0$ satisfies the  quadratic equation
\begin{equation}
a_3 y^{2}_0-(2a_1 +a_2)y_0 -6=0.
\label{eq:59}
\end{equation}
Therefore  there are two branches corresponding to $(-1,y_{0j}),
~j=1,2$. The resonances satisfy  equation (\ref{eq:31}a). Now one
should determine $y_{0j}$ and $a_i$, $i=1,2,3$ such that one of
the branches is the principal branch.

If $y_{0j}$ are the roots of (\ref{eq:59}), by setting
\begin{equation}
P(y_{0j})=-\left[a_3 y^{2}_{0j}-2(2a_1 +a_2)y_{0j} -18\right], \qquad j=1,2,
\label{eq:60}
\end{equation}
and if $(r_{j1},r_{j2})$ are the resonances corresponding to
$y_{0j}$, then one has
\begin{equation}
r_{j1}r_{j2}=P(y_{0j})=p_j, \qquad j=1,2,
\label{eq:61}
\end{equation}
where $p_j$
are integers  such that at least one is positive. Equation (\ref{eq:59})
gives that
\begin{equation}
a_3 =-\frac{6}{y_{01}y_{02}}, \qquad 2a_1 +a_2=a_3
(y_{01}+y_{02}).
\label{eq:62}
\end{equation}
Then (\ref{eq:60}) can be written as
\begin{equation}
P(y_{01})=6\left(1-\frac{y_{01}}{y_{02}}\right), \qquad P(y_{02})=6\left(1-
\frac{y_{02}}{y_{01}}\right).
\label{eq:63}
\end{equation}
For $p_1 p_2 \neq 0$, the $p_j$ satisfy the following Diophantine equation
\begin{equation}
\frac{1}{p_1}+\frac{1}{p_{2}}=\frac{1}{6}.
\label{eq:64}
\end{equation}
Now one should determine all finite integer solutions  of  Diophantine equation.
One solution of (\ref{eq:64}) is $(p_1 , p_2)=(12,12)$. The following cases
should be considered: {\bf i)} If $p_1 >0$, $p_2 >0$ and $p_1 < p_2 $, then $p_1
>6$ and $p_2 >12$. {\bf ii)} If $p_1 >0$, $p_2 <0$, then $p_1 <6$. Based on these
observations there are the following nine integer solutions  of Diophantine
equation, viz
\begin{gather}
(p_{1},p_{2})=(12,12), (7,42), (8,24), (9,18), (10,15),\nonumber\\
\phantom{(p_{1},p_{2})={}}{} (2,-3), (3,-6),  (4,-12), (5,-30).
\label{eq:65}
\end{gather}
For each  $(p_1 , p_2 )$, one
should write $(r_{j1},r_{j2})$ such that $r_{ji}$ are distinct integers and
$r_{j1}r_{j2}=p_j$, $j=1,2$. Then  $y_{0j}$ and $a_i$ can be obtained
from (\ref{eq:62}) and (\ref{eq:63})  and
\begin{equation}
r_{j1}+r_{j2}=a_{1}y_{0j}+7, \qquad j=1,2,
\label{eq:66}
\end{equation}
 There are the five following cases such that all the resonances are distinct
integers for both branches. The resonances and the simplified equations for these cases are
 \begin{align}
\mbox{ II.a:~} & a_{1}=a_{2}=0, \ y^{2}_{01}=\ds\frac{6}{a_3}:
(r_{11},r_{12})=(3,4), \quad
y_{02}=-y_{01}: \ (r_{21},r_{22})=(3,4),\nonumber\\
 & y'''=a_{3}y^{2}y', \label{eq:67} \\
\mbox{ II.b:~} &  y_{01}=-\frac{1}{a_1}: \ (r_{11},r_{12})=(2,4), \quad
y_{02}=\frac{3}{a_1}: \ (r_{21},r_{22})=(4,6),\nonumber\\
 & y'''=a_{1}\left(yy''+2y'{}^{2}+2a_{1}y^{2}y'\right),
\label{eq:68} \end{align}
\begin{align} \mbox{ II.c:~} &
y_{01}=-\frac{3}{a_1}: \ (r_{11},r_{12})=(1,3), \quad
y_{02}=-\frac{6}{a_1}: \
(r_{21},r_{22})=(-2,3),\nonumber\\
& y'''=a_{1}\left(yy''+y'{}^{2}-\frac{1}{3}a_{1}y^{2}y'\right),
\label{eq:68.a} \\
\mbox{ II.d:~} &  y_{01}=-\frac{2}{a_1}: \ (r_{11},r_{12})=(1,4),\quad y_{02}=-
\frac{6}{a_1}: \ (r_{21},r_{22})=(-3,4),\nonumber\\
&  y'''=a_{1}\left(yy''+2y'{}^{2}-\frac{1}{2}a_{1}y^{2}y'\right),
\label{eq:69}\\
\mbox{ II.e:~} &  y_{01}=-\frac{1}{a_1}: \ (r_{11},r_{12})=(1,5),\quad
y_{02}=-\frac{6}{a_1}: \ (r_{21},r_{22})=(-5,6),\nonumber\\
&  y'''=a_{1}\left(yy''+5y'{}^{2}-a_{1}y^{2}y'\right).
\label{eq:70}
\end{align}

For each case the compatibility conditions for the simplified equations are identically satisfied.
To find the canonical form of the third order equations of Painlev\'{e}
type, one should add nondominant terms with the coefficients which are
analytic functions of $z$.

{\bf  II.a:} By using the linear transformation (\ref{eq:36}) one can set
$2A_1 +A_3 =0$, $A_2 =0$ and $a_{3}=6$.
The compatibility conditions at $j=3,4,$  for the both branches
allow one to determine the coefficients $A_{k}$.
The canonical form of the equation for this case is
\begin{gather}
y'''=6 y^{2}y'-\left(\frac{1}{2}c^{2}_{1}z^2 -c_2 z -c_3\right)y'+c_1 y^2\nonumber\\
\phantom{y'''=}{} -\left(c^{2}_{1}z-c_2\right)y
-\frac{1}{4}c^{3}_{1}z^2 +\frac{1}{2}c_1 c_2 z +\frac{1}{2}c_1 c_3,
\label{eq:74}
\end{gather}
where $c_1$, $c_2$, $c_3$ are constants.
If one replaces  $z-\frac{c_2}{c^{2}_1}$ by $z$, $y$ by
$\gamma y$ and $z$ by $\delta z$ such that
$\gamma \delta =1$ and $ c_1 \delta ^2=-2$, then (\ref{eq:74}) yields
\begin{equation}
u'''=6 u^{2}u'+12zuu' +4\left(z^2 +k\right)u'+4zu+4u^2,
\label{eq:77}
\end{equation}
where $u=y-z$ and $k$ is a constant.
Equation (\ref{eq:77}) was  considered in~\cite{cazy, bur3}, and its first integral
is PIV.

If $c_1 =c_2 =0$, (\ref{eq:74})
can be solved in terms of  elliptic functions.
If $c_1 =0$ and $c_2 \neq0$, (\ref{eq:74}) gives
\begin{equation}
y'''=6 y^{2}y' +c_2\left(z+\frac{c_3}{c_2}\right) y'+c_2 y.
\label{eq:79}
\end{equation}
If one introduces $t=z+\frac{c_3}{c_2}$, then the first integral of
(\ref{eq:79}) is PII.

{\bf II.b:} By using the linear transformation (\ref{eq:36})
one can always choose
$2A_1 +A_3 =0$, $A_2 =0$ and $a_{1}=-1$.
Then the  compatibility conditions for the both branches, that is the
arbitrariness of $y_{21}$ and
$y_{41} $ for the first branch and $y_{42}$ and $y_{62}$
for the second branch, imply that all
the coefficients $A_k$ of nondominant terms,  are zero.
So the canonical form for this case is
\begin{equation}
y'''=- yy'' -2y'{}^2+2 y^2 y'.
\label{eq:80}
\end{equation}
Equation (\ref{eq:80}) was  given in \cite{cazy, bur3}.

{\bf II.c:} By using the linear transformation (\ref{eq:36}) one can always set
$A_3 =A_5 =0$ and $a_{1}=-3$.
Then the compatibility conditions at $j=1,3$ give that
$A_1 =c_{1}/2$, $A_2 =c_1, c_{1}=\mbox{constant}$ and $A'_{4}=A_6$.
Then the canonical form of the equation is
\begin{equation}
y'''=-3 yy''-3y'{}^{2}-3y^{2}y' +\frac{1}{2}c_{1}y''+c_1 yy'
+A_{4}y'+A'_{4}y+A_{7}.
\label{eq:82}
\end{equation}
The first integral of (\ref{eq:82}) gives that
\begin{equation}
u''=-3uu'-u^{3}+B_{1}u+B_{2},
\label{eq:83}
\end{equation}
where $u=y-(c_{1}/6)$,  and $ B_{1}$ and $B_{2}$ are arbitrary
analytic functions of $z$.
Equation~(\ref{eq:82}) was  considered in \cite{cazy}.

{\bf II.d:} One can always choose $A_3 =A_5 =0$ and $a_{1}=-2$
by the linear transformation~(\ref{eq:36}). The arbitrariness of
$y_{11}$ and $y_{41}$ for the
first branch and $y_{42}$ for the second branch imply that
$A_1 =A_2 =A_7= 0$ and $A'_4 =2A_6$.
The canonical form is
\begin{equation}
y'''=-2 yy''-4y'{}^{2}-2y^{2}y' +A_{4}y'+\frac{1}{2}A'_{4}y.
\label{eq:86}
\end{equation}
The first integral of (\ref{eq:86}) is
\begin{equation}
y''=\frac{y'{}^{2}}{2y}-2yy'-\frac{y^{3}}{2} +A_{4}y+\frac{c}{y},\qquad
c=\mbox{constant.}
\label{eq:87}
\end{equation}
The equation (\ref{eq:86}) was  considered in \cite{cazy, bur3}.

{\bf II.e:} By the linear transformation (\ref{eq:36}) one can choose
$A_1 =A_3 =0$ and $a_{1}=-1$.
Then the compatibility conditions give that
$A_2 =A_5 = 0$,  $A_6=A'_4 /3$ and $A_7=-A_{4}'' /3$.
After the  replacement of $y$ by $-y$ and $A_4$ by $3A_4$ the canonical form of
the equation for this case is
\begin{equation}
y'''=yy''+5y'{}^{2}-y^{2}y' +3A_{4}y'+A'_{4}y+{A_{4}''}.
\label{eq:89}
\end{equation}
Equation  (\ref{eq:89}) has the first integral
\begin{gather}
\left(y''-yy'-y^3 +A_4 y +A'_4\right)^2 =\frac{8}{3}\left(y'-y^2 \right)\left(y' +\frac{y^2}{2}+
\frac{3}{2}A_{4}\right)\nonumber\\
\qquad {}+4\left(y'-y^2\right )\left(2A_4 y^3 +A'_4 y+A''_4\right)+4A^2_4 y^2+4A_4A'_4 y
+4A'_4{}^2 +c,
\label{eq:90}
\end{gather}
where $A_4$ is an arbitrary function of $z$  and $ c$ is an
arbitrary  constant of integration.
Equation (\ref{eq:89}) was also considered in~\cite{cazy, bur3}.

{\bf Case III.} $a_4 \neq 0$: In this case there are three
branches corresponding to $(-1, y_{0j})$, $j=1,2,3$, where
$y_{0j}$ are the roots of (\ref{eq:31}b). Equation (\ref{eq:31}b)
implies that
\begin{equation}
\sum_{j=1}^{4} y_{0j}=\frac{a_3}{a_4},\qquad
\sum_{i \neq j} y_{0i}y_{0j}=\frac{1}{a_4}(2a_1 +a_2),\qquad
\prod_{j=1}^{3} y_{0j}=-\frac{6}{a_4}.
\label{eq:91}
\end{equation}
If the resonances (except $r_0 =-1$, which is common for all branches)
are $r_{ji}$, $i=1,2$,
corresponding to $y_{0j}$ and if one sets
\begin{equation}
P(y_{0j})=-\left[a_3 y^{2}_{0j}-2(2a_1 +a_2)y_{0j} -18\right],\qquad j=1,2,3,
\label{eq:92}
\end{equation}
then (\ref{eq:31}a) implies that
\begin{equation}
\prod_{i=1}^{2} r_{ji}=P(y_{0j})=p_j,
\label{eq:93}
\end{equation}
where $p_j$ are integers and in order to have a principal branch
at least one of them is
positive. Equations (\ref{eq:91}) and (\ref{eq:92}) give
\begin{gather}
p_1 =6\left(1-\frac{y_{01}}{y_{02}}\right)\left(1-\frac{y_{01}}{y_{03}}\right),\qquad
p_2 =6\left(1-\frac{y_{02}}{y_{01}}\right)\left(1-\frac{y_{02}}{y_{03}}\right),\nonumber\\
p_3 =6\left(1-\frac{y_{03}}{y_{01}}\right)\left(1-\frac{y_{03}}{y_{02}}\right)
\label{eq:94}
\end{gather}
and hence, the $p_j$ satisfy the following Diophantine  equation
\begin{equation}
\sum_{j=1}^3\frac{1}{p_j} =\frac{1}{6}.
\label{eq:95}
\end{equation}
Moreover equation (\ref{eq:94}) gives that
\begin{equation}
\prod_{j=1}^{3} p_{j}=-\frac{6^3}{(y_{01}y_{02}y_{03})^2}
(y_{01}-y_{02})^{2}(y_{01}-y_{03})^{2}
(y_{02}-y_{03})^{2},
\label{eq:96}
\end{equation}
if $a_{1} \neq 0$, that is, if $p_1 >0$, then either $p_2$ or $p_3$
is a negative integer. One should consider the
case $a_1 =0$ separately.

{\bf III.a:}  $a_1 =0$: In this case the sum of the resonances
for all three branches
is fixed and
\begin{equation}
\sum_{i=1}^2 r_{ji} =7,\qquad j=1,2,3.
\label{eq:97}
\end{equation}
Under this condition the solutions of the Diophantine equation
(\ref{eq:95}) are
$(p_1 , p_2 , p_3 )=(10, 10, -30)$ and $(10, 12, -60)$.

{\bf III.a.i:} $(p_1 , p_2 , p_3 )=(10, 10, -30)$: Equation
(\ref{eq:94}) can be written as
\begin{equation}
p_1 (y_{02}-y_{03})=ky_{01},\qquad p_2 (y_{03}-y_{01})=ky_{02},
\qquad p_3 (y_{01}-y_{02})=ky_{03},
\label{eq:99}
\end{equation}
where
\begin{equation}
k=\frac{6}{y_{01}y_{02}y_{03}} (y_{01}-y_{02})(y_{02}-y_{03})(y_{01}-y_{03}).
\label{eq:100}
\end{equation}
For $k=\pm 10 \sqrt{5}$
the system (\ref{eq:99}) has the nontrivial solutions $y_{0j}$,
$j=1,2,3$. For these values of  $y_{0j}$
the resonances and the coefficients $a_i$, $i=2,3,4$, are
\begin{gather}
y_{01}=\nu\left(1-\sqrt{5}\right):\ (r_{11}, r_{12})=(2,5),\quad
y_{02}=\nu\left(1+\sqrt{5}\right): \ (r_{21}, r_{22})=(2,5),\nonumber\\
y_{03}=6\nu: \ (r_{31}, r_{32})=(-3,10),\nonumber\\
a_2 =\frac{2}{\nu} , \quad a_3 =\frac{2}{\nu ^{2}} ,
\quad a_4 =\frac{1}{4\nu ^{3}}, \quad \nu=\mbox{constant},
\label{eq:102}
\end{gather}
for both values of $k$. For these values of $y_{0j}$ and $a_i$
the simplified equation passes the
Painlev\'{e} test for all branches.
The linear transformation (\ref{eq:36}) and the compatibility
conditions at the resonances of the first
and second branches are enough to determine all  coefficients
$A_{k}(z)$ of the nondominant terms.
The canonical form of the equation for this case is,
\begin{equation}
y'''=12y'{}^{2}+72y^2 y'+54y^4 +c_1,
\label{eq:103}
\end{equation}
where $c_1$ is an arbitrary constant. Equation (\ref{eq:103}) can be
obtained with the choice of
$\nu =1/\left(1-\sqrt{5}\right)$ and replacement of $y$ with $6y/\left(1-\sqrt{5}\right)$.
Equation (\ref{eq:103}) was  given in~\cite{cazy, cos1}.

{\bf III.a.ii:} $(p_1 , p_2 , p_3 )=(10, 12, -60)$:
For this case  equation (\ref{eq:99})
has nontrivial solution $y_{0j}$ for
$k=\pm 20 \sqrt{3}$. Then $y_{0j}$, $a_{i}$ and the
corresponding resonances are as follows:
\begin{gather}
y_{01}=-\frac{1}{\nu}\left(-1 \pm \sqrt{3}\right):\ (r_{11}, r_{12})=(2,5),
\quad y_{02}=\pm \frac{\sqrt{3}}{\nu}: \ (r_{21}, r_{22})=(3,4),\nonumber\\
y_{03}=-\frac{1}{\nu}\left(-6 \pm \sqrt{3}\right): \ (r_{31}, r_{32})=(-5,12),\nonumber\\
a_2 =3  \frac{7 \pm 3\sqrt{3}}{11}\, \nu ,  \quad
a_3 = \frac{40 \pm 14\sqrt{3}}{11}\,{\nu}^{2} ,\quad
a_4 = \frac{7 \pm 3\sqrt{3}}{11}\, {\nu}^{3},\quad \nu=\mbox{constant}.\!\!
\label{eq:104}
\end{gather}
By using the linear transformation (\ref{eq:36}) one can
choose $\nu =\pm \sqrt{3}$ and
$A_{1}=A_{2}=0$.
All other coefficients $A_{k}$ of the nondominant
terms can be determined from the  compatibility
conditions at the resonances of the first and second branches.
The canonical form for this case is as follows:
\begin{gather}
y'''=\frac{27 \pm 21 \sqrt{3}}{11} \left(y'{}^{2}+y^4 \right)
+\frac{120 \pm 42 \sqrt{3}}{11} y^2 y'
\nonumber\\
\phantom{y'''=}{}+c\left(\pm \frac{1 \mp \sqrt{3}}{\sqrt{3}} \, y'+y^2 \right)
-\frac{231 \pm 143 \sqrt{3}}{132} c^2
\label{eq:105}
\end{gather}
or
\begin{gather}
y'''=6y^{2}y'+\frac{3}{11}\left(9 \pm 7 \sqrt{3}\right)\left(y'+y^2 \right)^{2}
-\frac{1}{22}\left(4 \mp 3 \sqrt{3}\right)c_{1}y'
\nonumber\\
\phantom{y'''=}{}+\frac{1}{44}\left(3 \mp 5 \sqrt{3}\right)c_{1}y^2 -\frac{1}{352}\left(9 \pm 7 \sqrt{3}\right)c_{1}^{2},
\label{eq:106}
\end{gather}
where $c_{1}=44c/\left(3 \mp 5 \sqrt{3}\right)$. Equation  (\ref{eq:106}) was
considered in~\cite{cos1}.

{\bf III.b:}  $a_1 \neq 0$: Since  $p_1 ,p_2 >0$, $p_3 <0$, equation
(\ref{eq:96}) can be written as
\begin{equation}
p_{1}p_{2}\hat{p}_{3}=6n^2,
\label{eq:107}
\end{equation}
where $n$ is a constant and $\hat{p}_{3}=-p_3 $.
Then the Diophantine equation (\ref{eq:95}) yields
\begin{equation}
p_{1}p_{2}=\hat{p}_{3}(p_{1}+p_{2})-n^2
\label{eq:108}
\end{equation}
and, since $(p_{1}-p_{2})^2 \geq 0$, then
\begin{equation}
(p_{1}+p_{2})^{2}-4\hat{p}_{3}(p_{1}+p_{2})+4n^2 \geq 0.
\label{eq:109}
\end{equation}
Therefore
$0<\hat{p}_{3}\leq n$.
Hence one may assume that $n$ is a positive integer.
When $\hat{p}_{3}=n$, equations (\ref{eq:107})
and (\ref{eq:108}) give  $(p_1 ,p_2 ,p_3 )=(6,n,-n)$
as the solution of the Diophantine equation.
For the case of $\hat{p}_{3}<n$, if one assumes that
$p_{1}<p_{2}$ (if $p_{1}=p_{2}$,
(\ref{eq:108}) implies that
$p_{1}$ and $p_{2}$ are complex numbers), then the Diophantine
equation (\ref{eq:95}) implies that
$p_{1}<12$. Equations  (\ref{eq:95}) and  (\ref{eq:108}) give that
\begin{equation}
(p_{1}{\hat p}_{3})^{2}=n^2[6p_{1}-(6-p_{1}){\hat p}_{3}],\qquad
(p_{1}p_{2})^{2}=n^2[6p_{1}+(6-p_{1})p_{2}]
\label{eq:110}
\end{equation}
for $p_{1}<6$ and for $6<p_{1}<12$ respectively. Equation~(\ref{eq:110}) imply that
$[6p_{1}-(6-p_{1}){\hat p}_{3}]$ and
$[6p_{1}+(6-p_{1})p_{2}]$ must be squares of  integers.
By the use of these
results, $p_{j}$ the multiplication of the resonances
for the branches  corresponding the
$y_{0j}$, $j=1,2,3$, are
\begin{gather}
(p_{1},p_{2},p_{3})=(4,6,-10), (5,870,-26), (5,195,-21),
(7,41,-1722), \nonumber\\
\phantom{(p_{1},p_{2},p_{3})={}}(7,38,-399), (7,33,-154), (8,22,-264), (8,16,-48), \nonumber\\
\phantom{(p_{1},p_{2},p_{3})={}}(9,15,-90), (10,14,-210), (11,13,-858).
\label{eq:111}
\end{gather}
For each values of $(p_{1}, p_{2}, p_{3})$ given in
(\ref{eq:111}) one should follow the
given steps below for $(4,6,-10)$.

When $(p_{1}, p_{2}, p_{3})=(4,6,-10)$, $p_{1}=4$ implies
that  possible integral values of $r_{1i}$, $i=1,2$ are
$(r_{11}, r_{12})=(1,4), (-1,-4)$. Then
\begin{equation}
r_{j1}+r_{j2}=a_{1}y_{0j}+7,\qquad j=1,2,3,
\label{eq:112}
\end{equation}
implies that $y_{01}=-2/a_{1}$ and $y_{01}=-12/a_{1}$ for
$(r_{11}, r_{12})=(1,4), (-1,-4)$ respectively.
On the other hand $y_{0j}$ satisfies  equation (\ref{eq:99}) for  $k=\pm 20$.
For $k=20$, $y_{02}=-9y_{01}/14$, but the resonance equation for the second
branch
\begin{equation}
r^{2}_{2i}-(7+a_{1}y_{02})r_{2i}+p_{2}=0,\qquad i=1,2,
\label{eq:113}
\end{equation}
implies that $7+a_{1}y_{02}$  be an integer.
So, in order to have integer resonances $(r_{21},r_{22})$
 for the second branch, $a_{1}y_{02}$ has to be integral.
 A similar argument holds for the
third branch, but for $k=20$, both $y_{01}$ and $y_{02}$
are not integers. Also for $k=-20$ the resonances
for the second and third branches are not integers.
Following the same steps one cannot find the integral
resonances   for the second and third branches for all
other cases of $(p_{1},p_{2},p_{3})$ given in~(\ref{eq:111}).

When $(p_{1},p_{2},p_{3})=(6,n,-n)$,  equation (\ref{eq:99})
has a nontrivial solution $y_{0j}$ for
$k=\pm n$. We have $y_{01}=0$, $y_{02}=y_{03}$ for $k=n$ and
$y_{01}=12\nu$, $y_{02}=\nu (6-n)$,
$y_{03}=\nu (6+n)$ for $k=-n$, where $\nu$ is an arbitrary constant.
Since $y_{01}=0$ for $k=n$,
this case is not be considered. For $k=-n$,  $2a_{1}+a_{2}$,
$a_{3}$ and $a_{4}$ can be determined
from  equation (\ref{eq:91}) to be
\begin{equation}
2a_{1}+a_{2}=-\frac{180-n^{2}}{2\nu (36-n^{2})},
\qquad a_{3}=-\frac{12}{{\nu}^{2} (36-n^{2})},\qquad
a_{4}=-\frac{1}{2{\nu}^{3} (36-n^{2})}.
\label{eq:114}
\end{equation}
Since $p_{1}=6$, then all possible distinct integral resonances
for the first branch are $(r_{11}, r_{12}) $ $=(-1,-6), \ (-2,-3),
\ (1,6),\ (2,3)$. Because of the double resonance,
$r_{0}=r_{11}=-1$, the case $(-1,-6)$ is not considered. When
$(r_{11},r_{12})=(1,6)$, equation (\ref{eq:112}) implies that
$a_{1}=0$. This case was considered in case III.a. For the other
possible resonances, $(-2,-3)$ and $(2,3)$, one can obtain the
$a_{i}$, $i=1,2,3,4,$ and $y_{0j}$, $j=1,2,3$. Once the
coefficients of the resonance equation (\ref{eq:31}a) are known
one should look at the distinct integer resonances for the second
and third branches. We have only two cases, such that all the
resonances are distinct integers for all branches. The resonances
and the corresponding simplified equations are as follows:
\begin{gather}
\mbox{III.b.i: \ \ }y_{01}=-\frac{12}{a_{1}}:\ (r_{11}, r_{12})=(-2,-3),\nonumber\\
\phantom{\mbox{III.b.i: \ \ }}y_{02}=-\frac{1}{a_{1}}(6-n): \ (r_{21}, r_{22})=(1,n),\nonumber\\
\phantom{\mbox{III.b.i: \ \ }}y_{03}=-\frac{1}{a_{1}}(6+n):\ (r_{31}, r_{32})=(1,-n),
\nonumber\\
y'''=a_{1}\!\left[yy''\!+\frac{3(12+n^{2})}{2(36-n^{2})} \,
y'{}^{2}\!- \frac{12}{36-n^{2}}\, a_{1}y^{2}y'\!
+\frac{1}{2(36-n^{2})}\, a^{2}_{1}y^{4}\right],\!\quad n\neq
1,6.\!\!\! \label{eq:115}
\end{gather}
It should be noted that as $n \rightarrow \infty$ the
simplified equation reduces~to (\ref{eq:58a}).
\begin{gather}
\mbox{III.b.ii:\ \ }y_{01}=-\frac{2}{a_{1}}: \ (r_{11}, r_{12})=(2,3),\nonumber\\
\phantom{\mbox{III.b.ii:\ \ }}y_{02}=-\frac{1}{a_{1}}\left(1-\frac{n}{6}\right):\ (r_{21}, r_{22})=(6,n/6),
\nonumber\\
\phantom{\mbox{III.b.ii:\ \ }}y_{03}=-\frac{1}{a_{1}}\left(1+\frac{n}{6}\right):\ (r_{31}, r_{32})=(6,-n/6),
\nonumber\\
y'''=a_{1}\left[yy''+\frac{468-n^{2}}{36-n^{2}}\, y'{}^{2}-
\frac{432}{36-n^{2}} \, a_{1}y^{2}y'+
\frac{108}{36-n^{2}}\, a^{2}_{1}y^{4}\right], \quad n\neq 6,36.
\label{eq:116}
\end{gather}

The canonical form of the equations corresponding to the
above cases can be obtained by adding the
nondominant terms with the analytic coefficients $A_{k}$, $k=1,\ldots,7$.

{\bf III.b.i:} By using the transformation (\ref{eq:36})
one can set $A_{3}=A_{4}=0$ and
$a_{1}=2$. The compatibility conditions at the resonances
imply that all the coefficients are zero
except $A_{6}$ and $A_{7}$ which remain arbitrary for $n=2$.
For $n=3$ $A_{7}$ is arbitrary
and all the other coefficients are zero. For $n=4,5$ and $6$  all
 coefficients $A_{k}$ are zero.
However, it was proved in~\cite{pic} that for $n \geq 4$ the
equation does not admit
nondominant terms. The canonical forms of the equations for
$n=2$ and $n=3$ are
\begin{gather}
y'''=2yy''+\frac{3}{2}y'{}^{2}-\frac{3}{2}{y}^{2}y'
+\frac{1}{8}{y}^{4}+A_{6}y+A_{7},
\label{eq:117}\\
y'''=2yy''+\frac{7}{3} y'{}^{2}-\frac{16}{9}{y}^{2}y'
+\frac{4}{27}{y}^{4}+A_{7},
\label{eq:118}
\end{gather}
respectively. Equations~(\ref{eq:117}) and~(\ref{eq:118}) were
given in~\cite{cazy} and~\cite{cos1},
and both can be linearized by letting $y=-2u'/u$ and
$y=-3u'/2u$ respectively.

{\bf III.b.ii:} The linear transformation (\ref{eq:36}) and
the compatibility conditions at the resonances of the first
and second branches
give the  canonical form as
\begin{gather}
y'''=-2yy''+\frac{26-2m^{2}}{m^{2}-1}y'{}^{2}
+\frac{24}{m^{2}-1}\left(2y'+y^{2}\right)y^{2}
+A_{5}\left(y'+y^2 \right)\nonumber\\
\phantom{y'''=}{}
-\frac{m^{2}-1}{48}\left(A_{5}''-\frac{1}{2}A_{5}^{2}\right)+c_1 z+c_2,
\label{eq:121}
\end{gather}
where $m=6/n$, $m\neq 1,6$,  $c_1$ and $c_2$ are arbitrary
constants and $A_5$ is an arbitrary
function of~$z$. Equation (\ref{eq:121}) was  given in~\cite{cazy} and~\cite{cos1} and is
equivalent to
\begin{equation}
y'+y^{2}=\frac{m^{2}-1}{48}A_{5}-\frac{m^{2}-1}{4}u,
\qquad u''=6u^{2}+\frac{1}{4(m^{2}-1)}(c_1 z+c_2 ).
\label{eq:122}
\end{equation}

\section{Fourth order equations: $\boldsymbol{\mbox{P}_{\rm II}^{(4)}}$}

Differentiation of (\ref{eq:29}) with respect to $z$ gives the terms
$y^{(4)}$, $yy'''$, $y'y''$, $y^{2}y''$, $yy'{}^{2}$, $y^{3}y'$, all of
which are of order $-5$ for $\alpha =-1$,
as $z \rightarrow z_0$.
Addition of  the term $y^5$, which is also of order $-5$, gives the
following simplified equation
\begin{equation}
y^{(4)}=a_{1}yy'''+a_{2}y'y''+a_{3}y^{2}y''+a_{4}yy'{}^{2}
+a_{5}y^{3}y'+a_{6}y^{5},
\label{eq:140}
\end{equation}
where $a_{i}$, $i=1,\ldots,6$, are constants. Substitution of
(\ref{eq:30}) into (\ref{eq:140})
gives the following equations for resonance $r$ and for
$y_0$, respectively,
\begin{subequations}
\begin{gather}
Q(r)=(r+1)\left\{r^{3}-(11+a_{1}y_{0})r^{2}-\left[a_{3}y_{0}^{2}
-(7a_{1}+a_{2})y_{0}-46\right]r\right.\nonumber\\
\left.\phantom{Q(r)=} {}-a_{5}y_{0}^{3}+2(2a_{3}+a_{4})y_{0}^{2}
-6(3a_{1}+a_{2})y_{0}-96 \right\}=0,\\
a_{6}y_{0}^{4}-a_{5}y_{0}^{3}+(2a_{3}
+a_{4})y_{0}^{2}-2(3a_{1}+a_{2})y_{0}-24=0.
\end{gather}\label{eq:141}
\end{subequations}
Equation (\ref{eq:141}b) implies that in general
there are  four branches
of Painlev\'{e} expansion, if $a_{6} \neq 0$, corresponding
to the roots $y_{0j}$, $j=1,2,3,4$.
Now one should determine $y_{0j}$ and
$a_{i}$ such that at least one of the branches
is the  principal branch.
Depending on the number of branches there are four cases.
Each case should be considered separately.

{\bf Case I.} $a_5 =a_6 =0$, $2a_3 +a_4 =0$: In this case
there is only one branch which should be the
principal branch. There are following two subcases which
will be considered separately.

{\bf I.a:} $a_{1}=0$: In this case the equation (\ref{eq:141}.a)
gives that the resonances
$(r_{1},r_{2},r_{3})$
(additional to $r_{0}=-1$) satisfy
$\sum\limits_{i=1}^{3}r_{i}=11$, $\prod\limits_{i=1}^{3} r_{i}=24$.
Under these conditions the
only possible distinct positive integer resonances are
$(r_{1},r_{2},r_{3})=(1,4,6)$. Then (\ref{eq:141})
implies that $a_3 =0$ and $y_{0}=-12/a_{2}$.
Therefore the simplified equation is
\begin{equation}
y^{(4)}=a_{2}y'y''.
\label{eq:142}
\end{equation}
To obtain the canonical form of the equation, one should add the
nondominant terms, viz~$y'''$, $yy''$, $y''$, $y'{}^{2}$,
$y^{2}y'$, $yy'$, $y'$, $y^{4}$, $y^{3}$, $y^{2}$, $y$, $1$, that
is  terms of order greater than~$-5$ as $z \rightarrow z_{0}$ with
coefficients $A_{k}(z)$, $k=1,\ldots,12,$ which are analytic
functions of~$z$. The coefficients $A_{k}$ can be determined by
using the linear transformation (\ref{eq:36}) and the
compatibility conditions at the resonances. One can choose
$a_{2}=-12$, $A_{2}=0$ and $2A_{3}-A_{6}+A_{9}=0$ by the linear
transformation (\ref{eq:36}). The compatibility conditions, that
is the arbitrariness of $y_{1}$, $y_{4}$, $y_{6}$, give that
\begin{gather}
A_{3}''-A_{3}^{2}=0,\quad
A_{1}'+A_{1}^{2}=A_{3}/3, \quad
A_{4}=6A_{1}, \quad A_{5}=A_{8}=A_{9}=0,\nonumber\\
A_{6}=2A_{3},
\quad A_{7}''-A_{3}A_{7}=2A_{1}A_{3}A_{1}'+2A_{1}^{2}A_{3}',\quad
A_{10}=A_{3}'-A_{1}A_{3},\nonumber\\
A_{11}=(A_{7}-A_{10})'-A_{1}(A_{7}-A_{10}),\quad
A_{12}'+A_{1}A_{12}=\frac{1}{6}(A_{7}-A_{10})^{2}.
\label{eq:145}
\end{gather}
According to the solution of (\ref{eq:145}a),
there are the  four following cases:

{\bf I.a.i:} $A_{3}=0$, $A_{1}=0$: Then the canonical
form of the equation is
\begin{equation}
y^{(4)}=-12y'y''+(c_{1}z+c_{2})y'+c_{1}y
+\frac{1}{18 c_{1}}(c_{1}z+c_{2})^{3}+c_{3},
\label{eq:147}
\end{equation}
where $c_{i}$, $i=1,2,3,$ are arbitrary constants.
Integration of~(\ref{eq:147}) once gives
\begin{equation}
y'''=-6y'{}^{2}+(c_{1}z+c_{2})y+\frac{1}{72 c_{1}^{2}}(c_{1}z+
c_{2})^{4}+c_{3}z+c_4,
\label{eq:148}
\end{equation}
where $c_{4}$ is an integration constant. If $c_{1} \neq 0$ and  $c_{2}=0$,
then the equation (\ref{eq:148}) takes the form of (\ref{eq:42}).
For $c_{1}=0$ and $c_{2} \neq 0$ (\ref{eq:148}) yields  (\ref{eq:41}).

{\bf I.a.ii:} $A_{3}=0$, $A_{1}=1/(z-c)$: Without loss
of generality one can choose the  constant of integration
$c$ as zero. Then
the canonical form of the equation is
\begin{gather}
y^{(4)}=-12y'y''+\frac{1}{z}y'''+\frac{6}{z}y'{}^{2}
+(c_{1}z-c_{2})y'+\frac{c_{2}}{z}y+
\frac{1}{24} c_{1}^{2}z^{3}\nonumber\\
\phantom{y^{(4)}=} {}-\frac{1}{9} c_{1}c_{2}z^{2}
+\frac{1}{12} c_{2}^{2}z+\frac{c_{3}}{z}.
\label{eq:152} \end{gather}
If $c_{1}=c_{2}=0$, then (\ref{eq:152}) is equivalent to
\begin{equation}
u'=\frac{1}{z}(u+c_{3}), \qquad y'''=-6y'{}^{2}+u.
\label{eq:153}
\end{equation}
If $c_{1}=0$ and $c_{2}\neq 0$,  after replacement of
$z$ by $\gamma z$,
$y$ by $\beta y$, such that $\beta \gamma=1$ and $c_{2}\gamma ^{3}=6$
equation (\ref{eq:152}) takes the form
\begin{equation}
y^{(4)}=-12y'y''+\frac{1}{z}y'''+\frac{6}{z}\left(y'{}^{2}+y\right)
-6y'+3z+\frac{\tilde{c}_{3}}{z},
\label{eq:154}
\end{equation}
where $\tilde{c}_{3}=c_{3}\gamma ^{4}$.
If $c_{1}\neq 0$ and $c_{2}=0$,
the equation (\ref{eq:152}) takes the form
\begin{equation}
y^{(4)}=-12y'y''+\frac{1}{z}\left(y'''+6y'{}^{2}\right)+12zy'
+6z^{3}+\frac{\tilde{c}_{3}}{z},
\label{eq:155}
\end{equation}
where $\tilde{c}_{3}$ is an arbitrary constant.

{\bf I.a.iii:} $A_{3}=6/(z-c)^{2}$: For simplicity let $c=0$.
Then the canonical form of the equation is
\begin{gather}
y^{(4)}=-12y'y''+A_{1}\left(y'''+6y'{}^{2}\right)+\frac{6}{z^2}
(y''+2yy')\nonumber\\
\phantom{y^{(4)}=}{}+A_{7}y'+A_{10}y^{2}+A_{11}y+A_{12},
\label{eq:156}
\end{gather}
where
\begin{gather}
A_{1}=\frac{2c_{1}z^{3}-c_{2}}{z\left(c_{1}z^{3}+c_{2}\right)},\nonumber\\
A_{7}=\frac{1}{c_{1}z^{3}+c_{2}}\left(\frac{1}{5}c_{1}c_{3}z^{6}
+\frac{1}{5}c_{2}c_{3}z^{3}+c_{1}c_{4}z-24c_{1}
+c_{2}c_{4}z^{-2}-6c_{2}z^{-3}\right), \nonumber\\
A_{10}=-\frac{12}{z^{3}}
-\frac{6}{c_{1}z^{3}+c_{2}}\left(2c_{1}-c_{2}z^{-3}\right),\nonumber
\end{gather}
\begin{gather}
A_{11}=\frac{1}{c_{1}z^{3}+c_{2}}\left(\frac{c_{1}c_{3}^{2}}{1350}z^{10}
+\frac{c_{2}c_{3}^{2}}{900}z^{7}+
\frac{c_{1}c_{3}c_{4}}{60}z^{5}+\frac{c_{2}c_{3}c_{5}}{15}z^{2}\right.\nonumber\\
\left.\phantom{A_{11}=}{}
+c_{5}z-\frac{c_{1}c_{4}^{2}}{6}
-\frac{c_{2}c_{4}^{2}}{24}z^{-3}\right),\nonumber\\
A_{12}=\frac{-1}{\left(c_{1}z^{4}+c_{2}z\right)^{2}}\left[c_{1}^{2}c_{3}z^{10}
-48c_{1}^{2}z^{8}
+\frac{4c_{1}c_{2}c_{3}}{5}z^{7}
+5c_{1}^{2}c_{4}z^{5}-\frac{c_{2}^{2}c_{3}}{5}z^{4}\right.\nonumber\\
\phantom{A_{12}=}{}+4c_{1}c_{2}c_{4}z^{2}
-42c_{1}c_{2}z-
c_{2}^{2}c_{3}z^{-1}+6c_{2}^{2}z^{-2}-\left(c_{1}z^{4}+c_{2}z\right)\nonumber\\
\left. \phantom{A_{12}=}{}\times
\left(\frac{6c_{1}c_{3}}{5}z^6
+\frac{3c_{2}c_{3}}{5}z^{3}-48c_{1}+c_{1}c_{4}
-2c_{2}c_{4}z^{-2}+6c_{2}z^{-3}\right)\right],
\label{eq:157} \end{gather}
where $c_{i}$,  $i=1,\ldots,5,$ are constants. Equations
(\ref{eq:147}), (\ref{eq:152}),
(\ref{eq:154}),
(\ref{eq:155}) and (\ref{eq:156}) were  considered
in \cite{bur3,cos2,marty}.

{\bf I.a.iv:} $A_{3}=6{\mathcal P}(z;0,\alpha_{1} )$:
 If one replaces  $A_3$ with $6 \hat A_{3}$ in the equation
(\ref{eq:145}a), then Weierstrass elliptic function
${\mathcal P}(z;0,\alpha_1)$,
where $\alpha_1$ is an arbitrary integration constant, is a~solution of
the resulting equation. If one lets
$A_{1}={\tilde A}'_{1}/{\tilde A}_{1}$,
then  the
equation (\ref{eq:145}b) gives   Lam{\'e}'s
equation for ${\tilde A}_{1}$.
Therefore,
\begin{equation}
{\tilde A}_{1}= c_1 E_1 (z) +c_2 F_1 (z),
\label{eq:155a}
\end{equation}
where $c_1$ and $c_2$ are constants and $E_1 (z)$ and
$F_1 (z)$ are the Lam{\' e}
functions of degree one and of
the first and second kind respectively, which were given in (\ref{eq:51}).
Similarly replacement of $A_1$ with $6 \hat A_{1}$, $A_3$
with $6 \hat A_{3}$ and $A_7$ with $6 \hat A_{7}$
in the equation for $A_7$ in (\ref{eq:145}) yields
\begin{equation}
{\hat A}_{7}''-6{\mathcal P}(z;0,
\alpha_{1}){\hat A}_{7}=72\left({\hat A}_{1}
{\hat A}_{3}{\hat A}_{1}'+{\hat A}_{1}^{2}{\hat A}_{3}'\right).
\label{eq:155b}
\end{equation}
So the homogenous solution of the above equation
is nothing but the Weierstrass
elliptic function $ {\mathcal P}(z;0,\alpha_{1})$.
Therefore the compatibility
conditions (\ref{eq:145}) allow one to determine all the nonzero
coefficients $A_{k}(z)$  in terms of the Weierstrass
elliptic function $ {\mathcal P}(z)$.

{\bf I.b:} $a_{1} \neq 0$: Equation (\ref{eq:141}a)
implies that $r_{1}r_{2}r_{3}=24$. Under this
condition there are four possible cases of
$(r_{1},r_{2},r_{3})$ such that $r_{i}>0$ and
distinct integers, but there is only the following
case out of the four cases such that the compatibility
conditions at the resonances for the simplified equations
are identically satisfied and $y_{0} \neq 0$
\begin{equation}
(r_{1},r_{2},r_{3})=(2,3,4),\qquad y_{0}=-2/a_{1},
\qquad a_{2}=3a_{1},\qquad a_{3}=a_{4}=0.
\label{eq:158}
\end{equation}
By adding the nondominant terms to the simplified equation,
using the
linear
transformation (\ref{eq:36}) and the compatibility conditions
one finds  the canonical form
of the equation
\begin{gather}
y^{(4)}=-2yy'''-6y'y''+A_{1}\left(y'''+2yy''+2y'{}^{2}\right)\nonumber\\
\phantom{y^{(4)}=}{}+A_{3}(y''+2yy')+A_{7}\left(y'+y^{2}\right)+A_{12},
\label{eq:159}
\end{gather}
where $A_{1}$, $A_{3}$, $A_{7}$ and $ A_{12}$ are arbitrary
functions of $z$. If one lets $u=y^{2}+y'$,
then the equation (\ref{eq:159}) can be linearized.
Equation (\ref{eq:159}) was  considered in \cite{bur3, cos2}.

{\bf Case II.} $a_5 =a_6 =0$: In  this case there are
two branches corresponding
to $(-1,y_{0j})$, $j=1,2$, where $y_{0j}$ are the
roots of (\ref{eq:141}.b) and
\begin{equation}
y_{01}+y_{02}=\frac{2(3a_{1}+a_{2})}{2a_{3}+a_{4}},
\qquad y_{01}y_{02}=-\frac{24}{2a_{3}+a_{4}}.
\label{eq:162}
\end{equation}
Let $(r_{j1},r_{j2},r_{j3})$ be the roots (additional to $r_{0}=-1$)
of the resonance equation~(\ref{eq:141}a) corresponding to $y_{0j}$. When one sets
\begin{equation}
P(y_{0j})=-2(2a_{3}+a_{4})y_{0j}^{2}+6(3a_{1}+a_{2})y_{0j}+96,\qquad j=1,2,
\label{eq:163}
\end{equation}
(\ref{eq:141}a) implies that
\begin{equation}
\prod _{i=1}^{3}r_{ji}=P(y_{0j})=p_{j}, \qquad j=1,2,
\label{eq:164}
\end{equation}
where the $p_{j}$ are integers and at least one of them is a
positive integer in order to have the principal branch.
Let the branch corresponding to $y_{01}$ be the
principal branch, that is $p_{1}>0$. Equations~(\ref{eq:162}) and (\ref{eq:163}) give
\begin{equation}
P(y_{01})=24\left(1-\frac{y_{01}}{y_{02}}\right)=p_{1},
\qquad P(y_{02})=24\left(1-\frac{y_{02}}{y_{01}}\right)=p_{2}.
\label{eq:165}
\end{equation}
Hence the $p_{j}$ satisfy the following Diophantine
equation, if $p_{1}p_{2} \neq 0$,
\begin{equation}
\frac{1}{p_{1}}+\frac{1}{p_{2}}=\frac{1}{24}.
\label{eq:166}
\end{equation}
There are 21 integer solutions $(p_{1},p_{2})$
of (\ref{eq:166}) such that one of the $p_j$
 is positive.
Once~$p_{1}$ is
known, for each $p_{1}$ one can write possible
distinct positive integers $(r_{11}, r_{12}, r_{13})$
such that $\prod\limits_{i=1}^{3} r_{1i}=p_1 $. Then for
each set of $(r_{11}, r_{12}, r_{13})$,  $a_{k}$,
$k=2,3,4,$ and $y_{0j}$ can be determined in terms of $a_{1}$ by using
\begin{equation}
\sum_{i=1}^{3} r_{ji}=11+a_{1}y_{0j},
\qquad \sum_{i \neq k} r_{ji}r_{jk}=-a_{3}y_{0j}^{2}+
(7a_{1}+a_{2})y_{0j}+46,
\label{eq:167}
\end{equation}
for $j=1$, and the equation (\ref{eq:162}).
Then for these values of $a_{k}$ and $y_{0j}$
one should find all  cases such that
the resonance equation (\ref{eq:141}a)
has  distinct integral roots $r_{2i}$
corresponding to $y_{02}$. Only for the following
cases {\bf a)} $(p_{1},p_{2})=(12,-24)$ and
{\bf b)} $(p_{1},p_{2})=(20,-120)$  are all the
resonances  distinct
integers for both branches,  one of which is the
principal branch. The resonances and the simplified
equations for these cases are as follows:
\begin{gather}
{\mbox{II.a:}\ \ }(p_{1},p_{2})=(12,-24):\nonumber\\
\phantom{{\mbox{II.a:}\ \ }}y_{01}=-\frac{3}{a_{1}}: \ (r_{11}, r_{12}, r_{13})=(1,3,4), \nonumber\\
\phantom{{\mbox{II.a:}\ \ }}y_{02}=-\frac{6}{a_{1}}: \ (r_{21}, r_{22}, r_{23})=(-2,3,4),\nonumber\\
\phantom{{\mbox{II.a:}\ \ }}y^{(4)}=a_{1}\left(yy'''+3y'y''
-\frac{1}{3}a_{1}y^{2}y''-\frac{2}{3}a_{1}yy'{}^{2}\right),
\label{eq:168}
\\
{\mbox{II.b:} \ \ }(p_{1},p_{2})=(20,-120):\nonumber\\
\phantom{{\mbox{II.b:} \ \ }}y_{01}=-\frac{1}{a_{1}}: \ (r_{11}, r_{12},
r_{13})=(1,4,5),\nonumber\\
\phantom{{\mbox{II.b:} \ \ }}y_{02}=-\frac{6}{a_{1}}: \ (r_{21}, r_{22}, r_{23})=(-5,4,6),\nonumber\\
\phantom{{\mbox{II.b:} \ \ }}y^{(4)}=a_{1}\left(yy'''+11y'y''
-a_{1}y^{2}y''-2a_{1}yy'{}^{2}\right).
\label{eq:169}
\end{gather}
For  case II.a the compatibility conditions at
the resonances of the simplified
equation are identically satisfied. For the case II.b
the compatibility condition at the
resonance $r_{13}=5$ implies that $y_{4}=0$ which contradicts
with the arbitrariness of $y_{4}$.
Moreover in the case~II.b, if one lets $y=\lambda u$ such
that $\lambda a_{1}=1$, integration
of the  simplified equation once yields
\begin{equation}
u'''=uu''+5u'{}^{2}-u^{2}u'+c,
\label{eq:170}
\end{equation}
where $c$ is an arbitrary integration constant. Equation
(\ref{eq:170}) is not a Painlev\'{e} type equation
unless $c=0$. It was  studied in~\cite{bur3, marty}.
Hence we  consider  case~II.a.
Adding the nondominant terms to the  simplified equation
and by using the linear transformation~(\ref{eq:36})
and the compatibility conditions of the first branch we can  determine the
coefficients $A_{k}(z)$ of the nondominant
terms. The canonical form of the
equation for the case~II.a is
\begin{gather}
y^{(4)}=-3yy'''-9y'y''-3y^{2}y''-6yy'{}^{2}+Ry''\nonumber\\
\phantom{y^{(4)}=}{}+2R'y'
+R''y+A_{9}\left(y^{3}+3yy'+y''-Ry\right)+A_{12},
\label{eq:171}
\end{gather}
where $R(z)=A_{3}(z)-A_{9}(z)$ and $A_{3}$ and $ A_{9}$ are
arbitrary analytic functions of $z$. If one lets
\begin{equation}
u=y''+3yy'+y^{3}-Ry,
\label{eq:172}
\end{equation}
then  equation (\ref{eq:171}) can be reduced to a
linear equation for $u$.
Equation (\ref{eq:171}) was  considered in \cite{bur3, cos2}.

{\bf Case III.} $a_6 =0$: There are three branches
corresponding to $y_{0j}$, $j=1,2,3,$ which are the roots of the
equation (\ref{eq:141}b). If one lets
\begin{gather}
\prod_{i=1}^{3} r_{ji}=p_{j}=P(y_{0j})=
a_{5}y_{0j}^{3}-2(2a_{3}+a_{4})y_{0j}^{2}\nonumber\\
\qquad \qquad \qquad {}+6(3a_{1} +a_{2})y_{0j}+96, \qquad j=1,2,3,
\label{eq:174}
\end{gather}
where $p_{j}$ are integers and at least one of them is
positive, by the use of the same procedure as was
carried in  the previous case the $p_{j}$  satisfy the
following Diophantine equation:
\begin{equation}
\sum _{j=1}^{3}\frac{1}{p_{j}}=\frac{1}{24},
\label{eq:175}
\end{equation}
if $p_{1}p_{2}p_{3} \neq 0$ and, if $a_1 \neq 0$,
\begin{equation}
\prod _{j=1}^{3} p_{j}=-\frac{24^{3}}{(y_{01}y_{02}y_{03})^{2}}
(y_{01}-y_{02})^{2}(y_{01}-y_{03})^{2}(y_{02}-y_{03})^{2}.
\label{eq:176}
\end{equation}
Let $p_{1},p_{2} >0$ and $p_{3} <0$. If $(r_{j1},
r_{j2},r_{j3})$ are the resonances corresponding to
$y_{0j}$ respectively, then they satisfy  equation
(\ref{eq:167}) for $j=1,2,3$.
There are the following  two cases which should be considered separately.

{\bf III.a:} $a_1 =0$: Equation (\ref{eq:167}a) for $j=1$
implies that there are five possible values of
$(r_{11},r_{12},r_{13})$ and hence five possible values
of $p_{1}$. For each value of $p_{1}$
one can solve (\ref{eq:175}) such that $p_{2}>0$, $p_{3}<0$
and both are integers. Then for each $(p_{1},p_{2},p_{3})$
the equations
\begin{gather}
p_{1}=24\left(1-\frac{y_{01}}{y_{02}}\right)\left(1-\frac{y_{01}}{y_{03}}\right),
\qquad p_{2}=24\left(1-\frac{y_{02}}{y_{01}}\right)\left(1-\frac{y_{02}}{y_{03}}\right),
\nonumber\\
p_{1}=24\left(1-\frac{y_{03}}{y_{01}}\right)\left(1-\frac{y_{03}}{y_{02}}\right),
\label{eq:178}
\end{gather}
give the  equations (\ref{eq:99}) for $y_{0j}$
for
\begin{equation}
k=\frac{24}{y_{01}y_{02}y_{03}} (y_{01}-y_{02})(y_{02}
-y_{03})(y_{01}-y_{03}).
\label{eq:180}
\end{equation}
The system (\ref{eq:99}) has nontrivial solution if
$k^{2}=-(p_{1}p_{2}+p_{1}p_{3}+p_{2}p_{3})$.
For each value of~$k$ one can find $y_{0j}$ and
$a_{i}$, $i=3,4,5,$ in terms of $a_{2}$.
Once the coefficients of the resonance equation
(\ref{eq:141}a) are known for all branches,
one should look at the cases such that the roots of
(\ref{eq:141}a) are distinct integers for the
second and third branches. There is only one case,
$(p_{1},p_{2},p_{3})=(40,40,-120)$, and $k=40 \sqrt{5}$.
The $y_{0j}$, the resonances  and
the simplified  equation for this case are as follows:
\begin{gather}
y_{01}=\frac{4}{a_{2}}\left(1-\sqrt{5}\right):\ (r_{11},r_{12},r_{13})=(2,4,5),\nonumber\\
y_{02}=\frac{4}{a_{2}}\left(1+\sqrt{5}\right):\ (r_{21},r_{22},r_{23})=(2,4,5),\nonumber\\
y_{03}=\frac{24}{a_{2}}: \ (r_{31},r_{32},r_{33})=(-3,4,10),\nonumber\\
y^{(4)}=a_{2}\left(y'y''+\frac{1}{8}a_{2}y^{2}y''
+\frac{1}{4}a_{2}yy'{}^{2}+\frac{1}{64}{a_{2}^{2}}y^3 y'\right).
\label{eq:181}
\end{gather}
The compatibility conditions are identically satisfied
for the simplified equation. To obtain the canonical form
of the equation one should add the nondominant terms with
analytic coefficients $A_{k}(z)$, $k=1,\ldots,12$.
The linear transformation (\ref{eq:36}) and the compatibility
conditions at the resonances of the first
and second branches give the following equation
\begin{equation}
y^{(4)}=24y'y''+72y^{2}y''+144yy'{}^{2}+216y^{3}y'{}^{2}.
\label{eq:182}
\end{equation}
Integration of (\ref{eq:182}) once gives (\ref{eq:103}).

{\bf III.b:} $a_1 \neq 0$: In this case the resonances
$(r_{j1},r_{j2},r_{j3})$ and $y_{0j}$
satisfy (\ref{eq:167}) for $j=1,2,3$  and
\begin{equation}
\sum _{i=1}^{3} y_{0j}=\frac{1}{a_{5}}(2a_{3}+a_{4}),\qquad
\sum_{j \neq k} y_{0j}y_{0k}=-\frac{2}{a_{5}}(3a_{1}+a_{2}),\qquad
\prod_{i=1}^{3} y_{0j}=-\frac{24}{a_{5}},
\label{eq:184}
\end{equation}
respectively.  The $p_{j}=\prod\limits_{j=1} ^{3}r_{ji}$ satisfy the
Diophantine equation (\ref{eq:175}).
If one lets
\begin{equation}
n^{2}=\frac{24^{2}}{(y_{01}y_{02}y_{03})^{2}}
(y_{01}-y_{02})^{2}(y_{01}-y_{03})^{2}(y_{02}-y_{03})^{2},
\label{eq:185}
\end{equation}
then (\ref{eq:176}) gives
\begin{equation}
p_{1}p_{2}\hat{p}_{3}=24n^{2},
\label{eq:186}
\end{equation}
where $\hat{p}_{3}=-p_{3}$ and
$p_{1}<48$.
If one follows the procedure given in the previous section,
(\ref{eq:175}) and  (\ref{eq:186}) give that
\begin{equation}
(p_{1}{\hat p}_{3})^{2}=n^2[24p_{1}-(24-p_{1}){\hat p}_{3}],\qquad
(p_{1}p_{2})^{2}=n^2[24p_{1}+(24-p_{1})p_{2}],
\label{eq:191}
\end{equation}
for $p_{1}<24$ and for $24<p_{1}<48$ respectively.
So the right hand sides of both equations in~(\ref{eq:191})
must be complete squares.
Based on these conditions on $p_{i}$, $i=1,2,3$, there are 71
integer solutions $(p_{1},p_{2},p_{3})$
of the Diophantine equation (\ref{eq:175}). For each solution
$(p_{1},p_{2},p_{3})$, one can find
$y_{0j}$ by solving the system of equations (\ref{eq:99}).
Then one can write possible resonances
$(r_{11},r_{12},r_{13})$ for each $p_{1}$ provided that
\begin{equation}
a_{1}y_{01}=\sum _{i=1}^{3} r_{1i}-11
\label{eq:192}
\end{equation}
are all integers. There are the  following three cases such
that all the resonances of all
three branches  are distinct integers.
\begin{gather}
\mbox{III.b.i:\ \ }(p_{1},p_{2},p_{3})=(15,60,-24):\nonumber\\
\phantom{\mbox{III.b.i:\ \ }}y_{01}=-\frac{2}{a_{1}}: \ (r_{11},r_{12},r_{13})=(1,3,5),\nonumber\\
\phantom{\mbox{III.b.i:\ \ }}y_{02}=-\frac{12}{a_{1}}:~(r_{21},r_{22},r_{23})=(-2,-5,6),\nonumber\\
\phantom{\mbox{III.b.i:\ \ }}y_{03}=-\frac{8}{a_{1}}:~(r_{31},r_{32},r_{33})=(-4,1,6),
\nonumber\\
\phantom{\mbox{III.b.i:\ \ }}a_{2}=\frac{11}{2}a_{1},\quad a_{3}=-\frac{1}{2}a_{1}^{2},
\quad a_{4}=-\frac{7}{4}a_{1}^{2},
\quad a_{5}=\frac{1}{8}a_{1}^{3}.
\label{eq:193}
\end{gather}
\begin{gather}
\mbox{III.b.ii:\ \ }(p_{1},p_{2},p_{3})=(24,n,-n), \quad n>0, \quad n \neq 24:\nonumber\\
\phantom{\mbox{III.b.ii:\ \ }}y_{01}=-\frac{2}{a_{1}}: \ (r_{11},r_{12},r_{13})=(2,3,4),\nonumber\\
\phantom{\mbox{III.b.ii:\ \ }}y_{02}=-\frac{1}{a_{1}}\left(1-\frac{n}{24}\right): \ (r_{21},r_{22},r_{23})=
\left(4,6,\frac{n}{24}\right),\nonumber\\
\phantom{\mbox{III.b.ii:\ \ }}y_{03}=-\frac{1}{a_{1}}\left(1+\frac{n}{24}\right):
 (r_{31},r_{32},r_{33})=\left(4,6,-\frac{n}{24}\right),\nonumber\\
\phantom{\mbox{III.b.ii:\ \ }} a_{2}=\frac{15552-3n^{2}}{576-n^{2}}a_{1},
\quad a_{3}=-\frac{6912}{576-n^{2}}a_{1}^{2},\nonumber\\
\phantom{\mbox{III.b.ii:\ \ }}a_{4}=-\frac{13824}{576-n^{2}}a_{1}^{2},\quad a_{5}=\frac{6912}{576-n^{2}}a_{1}^{3}.
\label{eq:194}
\end{gather}
\begin{gather}
\mbox{III.b.iii:\ \ }(p_{1},p_{2},p_{3})=(24,n,-n),\quad n>0,\quad n \neq 4,24:\nonumber\\
\phantom{\mbox{III.b.iii:\ \ }}y_{01}=-\frac{12}{a_{1}}: \ (r_{11},r_{12},r_{13})=(-2,-3,4),\nonumber\\
\phantom{\mbox{III.b.iii:\ \
}}y_{02}=-\frac{1}{a_{1}}\left(6-\frac{n}{4}\right): \
(r_{21},r_{22},r_{23})=\left(1,4,\frac{n}{4}\right),\nonumber
\\
\phantom{\mbox{III.b.iii:\ \
}}y_{03}=-\frac{1}{a_{1}}\left(6+\frac{n}{4}\right): \
(r_{31},r_{32},r_{33})=\left(1,4,-\frac{n}{4}\right),\nonumber
\end{gather}
\begin{gather}
\phantom{\mbox{III.b.iii:\ \ }}a_{2}=\frac{1152+2n^{2}}{576-n^{2}}a_{1},\quad a_{3}=-\frac{192}{576-n^{2}}a_{1}^{2},\nonumber\\
\phantom{\mbox{III.b.iii:\ \ }}a_{4}=-\frac{384}{576-n^{2}}a_{1}^{2}, \quad a_{5}=\frac{32}{576-n^{2}}a_{1}^{3}.
\label{eq:195}\end{gather}
For all three cases the simplified equations
pass the Painlev\'{e} test.
To obtain the canonical form of the equation
one should  add the nondominant terms
 with the coefficients $A_{k}(z)$, $k=1,\ldots,12$. The linear transformation~(\ref{eq:36})
and the compatibility conditions at the resonances
give the following equations:

{\bf III.b.i:}
\begin{gather}
y^{(4)}=-2yy'''-11y'y''-2y^{2}y''-7yy'{}^{2}-y^{3}y'{}^{2}+
A_{6}(y''+yy')\nonumber\\
\phantom{y^{(4)}=}{}+\frac{1}{3}A_{6}'\left(y^{2}+4y'\right)
+\frac{1}{3}A_{6}'''-\frac{2}{9}A_{6}A_{6}',
\label{eq:196}
\end{gather}
where $A_{6}$ is an arbitrary function of $z$. Equation (\ref{eq:196})
was  given in~\cite{cos2}.

{\bf III.b.ii:} The compatibility condition at the
resonance $r=6$ for the third branch
gives
\begin{equation}
A'_{1}+A_{1}^{2}=0.
\label{eq:197}
\end{equation}
So following two subcases should be considered separately.

{\bf III.b.ii.1:} $A_{1}=0$: The canonical form of the equation is
\begin{gather}
y^{(4)}=-2yy'''-\frac{6}{m^{2}-1}\left[\left(m^{2}-9\right)y'y''
-8y^{2}y''-16\left(yy'{}^{2}+y^{3}y'\right)\right]\nonumber\\
\phantom{y^{(4)}=}{}+A_{3}(y''+2yy')
+(A_{3}'+c_{1})\left(y'+y^{2}\right)+A_{12},
\label{eq:198}
\end{gather}
where $m=n/24$, $m \neq 1,4,6$, $A_{3}$ is an
arbitrary function of $z$ and
\begin{equation}
A_{12}=\frac{m^2 -1}{48}\left(A'''_{3}-A_{3}A'_{3}-
c_1 A_{3}+2c_{1}^{2}z+c_{2}\right),\qquad
c_{1},c_{2}=\mbox{constant}.
\label{eq:199}
\end{equation}
The result (\ref{eq:198}) was  given in~\cite{cos2}.

{\bf III.b.ii.2:} $A_{1}=1/(z-c)$: Without loss
of generality one can set $c=0$. The
canonical form of the equation is
\begin{gather}
y^{(4)}=-2yy'''+\frac{1}{m^{2}-1}\left[\left(54-6m^{2}\right)y'y''
+48y^{2}y''+96\left(yy'{}^{2}+y^{3}y'\right)\right]\nonumber\\
\phantom{y^{(4)}=}{}+\frac{1}{z}\left\{y'''+2yy''-\frac{1}{m^{2}-1}\left[\left(26-2m^{2}\right)y'{}^{2}
+48y^{2}y'+24y^{4}\right]\right\}\nonumber\\
\phantom{y^{(4)}=}{}+A_{3}(y''+2yy')+\left(A'_{3}-A_{3}\frac{1}{z}
+c_1 z\right)\left(y'+y^{2}\right)+A_{12},
\label{eq:200}
\end{gather}
where $A_{3}$ is an arbitrary function of $z$ and
\begin{equation}
A_{12}=-\frac{m^2 -1}{48}\left(A'''_{3}-\frac{1}{z}A''_{3}
-A_{3}A'_{3}+\frac{1}{2z}A_{3}^{2}
-c_{1}z A_{3}+\frac{1}{2}c_{1}^{2}z^3\right)+\frac{c_{2}}{z} ,
\label{eq:201}
\end{equation}
where $c_{1}$ and $c_{2}$ are constants. The result (\ref{eq:200}) was given in~\cite{cos2}.

{\bf III.b.iii:} If we let $m=n/4$, $m \neq 1,4,6,$
 the canonical form of the equation for $m=2$ is
\begin{gather}
y^{(4)}=2yy'''+5y'y''-\frac{3}{2}y^{2}y''-3yy'{}^{2}
+\frac{1}{2}y^{3}y'
+A_{1}\left[y'''-2yy''-\frac{3}{2}y'{}^{2}\right.\nonumber\\
\left.\phantom{y^{(4)}=}{}+\frac{3}{2}y^{2}y'
-\frac{1}{8}y^{4}-A_{7}y\right]
+A_{7}y'+A_{7}'y+A_{12}.
\label{eq:202}\end{gather}
If one sets
\begin{equation}
u=y'''-2yy''-\frac{3}{2}y'{}^{2}+\frac{3}{2}y^{2}y'-
\frac{1}{8}y^{4}-A_{7}y,
\label{eq:203}
\end{equation}
then (\ref{eq:202}) can be reduced to a linear equation for $u$.
It should be noted that (\ref{eq:203}) belongs to
$\mbox{P}_{\rm II}^{(3)}$ and was given in
(\ref{eq:117}). For $m=3$
\begin{gather}
y^{(4)}=2yy'''+\frac{20}{3}y'y''-\frac{16}{9}y^{2}y''
-\frac{32}{9}yy'{}^{2}+\frac{16}{27}y^{3}y'\nonumber\\
\phantom{y^{(4)}=}{}+A_{1}\left(y'''-2yy''-\frac{7}{3}y'{}^{2}
+\frac{16}{9}y^{2}y'-\frac{4}{27}y^{4}\right)+A_{12},
\label{eq:205}\end{gather}
where $A_{1}$ and $A_{12}$ are arbitrary functions of $z$.
If one sets
\begin{equation}
u=y'''-2yy''-\frac{7}{3}y'{}^{2}+\frac{16}{9}y^{2}y'-\frac{4}{27}y^{4},
\label{eq:206}
\end{equation}
(\ref{eq:205}) can be reduced to a linear equation in $u$.
Equation (\ref{eq:206}) belongs to $\mbox{P}_{\rm II}^{(3)}$
and was given in (\ref{eq:118}). Equations
(\ref{eq:202}) and (\ref{eq:205}) were  given
in \cite{cos2}. It should be noted, that for $m\geq 4$,
integration of the simplified equation once gives the
simplified equation of the case given in
(\ref{eq:116}) with an additional integration constant $c$.
Thus for $m\geq 4$ the simplified equation is not of
Painlev\'{e} type if $c\neq 0$.

{\bf Case IV.} $a_{6} \neq 0$: In this case there
are four branches corresponding to $(-1,y_{0j})$, $j=1,2,3,4$.
If $(r_{j1},r_{j2},r_{j3})$ are the resonances
corresponding to the branches,
$\prod\limits_{i=1}^{3} r_{ji}=p_{j}$  such that the $p_{j}$
are integers and at least one of them is positive. Then
(\ref{eq:141}a) implies that
\begin{equation}
P(y_{0j})=a_{5}y_{0j}^{3}-2(2a_{3}+a_{4})y_{0j}^{2}
+6(3a_{1}+a_{2})y_{0j}+96=p_{j},\qquad j=1,2,3,4.\!\!
\label{eq:208}
\end{equation}
On the other hand (\ref{eq:141}b) implies that
\begin{gather}
\sum _{j=1}^{4} y_{0j}=\frac{a_{5}}{a_{6}},
\qquad \sum _{j \neq i} y_{0j}y_{0i}=\frac{2a_{3}+a_{4}}{a_{6}},\nonumber\\
\sum _{j\neq i \neq k} y_{0j}y_{0i}y_{0k}=\frac{2(3a_{1}+a_{2})}{a_{6}},
\qquad \prod_{j=1}^{4}y_{0j}=-\frac{24}{a_{6}}.
\label{eq:209}
\end{gather}
Then (\ref{eq:208}) yields
\begin{equation}
p_{j}=P(y_{0j})=24\prod_{j \neq k}\left(1-\frac{y_{0j}}
{y_{0k}}\right),\qquad j=1,2,3,4.
\label{eq:210}
\end{equation}
Therefore the $p_{j}$ satisfy the Diophantine equation
\begin{equation}
\sum_{j=1}^{4} \frac{1}{p_{j}}=\frac{1}{24}.
\label{eq:211}
\end{equation}
To find the simplified equation one should follow
the following steps:
{\bf a)}~Find all integer solutions
$(p_{1},p_{2},p_{3},p_{4})$ of the Diophantine equation (\ref{eq:211}).
{\bf b)}~For each pair $(p_{1},p_{2})$
from the solution set of the Diophantine equation,
write all possible $(r_{j1},r_{j2},r_{j3})$ such that
$\prod\limits_{i=1}^{3} r_{ji}=p_{j}$, $j=1,2$.
{\bf c)}~Determine $y_{01}$ and $y_{02}$ in terms of $a_{1}$, if
$a_{1} \neq 0$, by using the equation (\ref{eq:167}a)
for $j=1,2$.
{\bf d)}~Use (\ref{eq:210}) to find $y_{03}$ and
$y_{04}$ in terms of $a_{1}$.
{\bf e)}~Eliminate the cases
$(r_{j1},r_{j2},r_{j3})$ $j=1,2,$ such that $a_{1}y_{0k}$, $k=3,4,$
are not integers (see the
equation~(\ref{eq:167}a)).
{\bf f)}~Find $a_{i}$, $i=2,\ldots,6,$ in terms of $a_{1}$ by
using the (\ref{eq:208}) and~(\ref{eq:209}). Once all the coefficients of the equation
(\ref{eq:141}a) are known, look at the cases such
that the roots of (\ref{eq:141}a) are distinct integers
for $ y_{03}$ and $y_{04}$.

There are four cases such that all the resonances are
distinct integers for all branches.
These cases and the
corresponding simplified equations are as follows:
\begin{gather}
\mbox{IV.a:\ \ } (p_{1},p_{2},p_{3},p_{4})=(6,-4,6,-24):\nonumber\\
\phantom{\mbox{IV.a:\ \ }}y_{01}=-\frac{5}{a_{1}}:\ (r_{11},r_{12},r_{13})=(1,2,3),\nonumber\\
\phantom{\mbox{IV.a:\ \ }}y_{02}=-\frac{10}{a_{1}}:\ (r_{21},r_{22},r_{23})=(-2,1,2),\nonumber\\
\phantom{\mbox{IV.a:\ \ }}y_{03}=-\frac{15}{a_{1}}:\ (r_{31},r_{32},r_{33})=(-3,-2,1),\nonumber\\
\phantom{\mbox{IV.a:\ \ }}y_{04}=-\frac{20}{a_{1}}:\ (r_{41},r_{42},r_{43})=(-4,-3,-2),\nonumber\\
\phantom{\mbox{IV.a:\ \ }}y^{(4)}=a_{1}\left(yy'''+2y'y''-\frac{2}{5}a_{1}y^{2}y''
-\frac{3}{5}a_{1}yy'{}^{2}
+\frac{2}{25}a_{1}^{2}y^{3}y'
-\frac{1}{625}a_{1}^{3}y^{5}\right),\!\!
\label{eq:213}\end{gather}
\begin{gather}
\mbox{IV.b:\ \ }(p_{1},p_{2},p_{3},p_{4})=(36,36,-84,-504):\nonumber\\
\phantom{\mbox{IV.b:\ \ }}y_{01}=-\frac{5}{a_{2}}:\  (r_{11},r_{12},r_{13})=(2,3,6),\nonumber\\
\phantom{\mbox{IV.b:\ \ }}y_{02}=\frac{10}{a_{2}}: \ (r_{21},r_{22},r_{23})=(2,3,6),\nonumber\\
\phantom{\mbox{IV.b:\ \ }}y_{03}=\frac{15}{a_{2}}: \ (r_{31},r_{32},r_{33})=(-2,6,7),\nonumber\\
\phantom{\mbox{IV.b:\ \ }}y_{04}=-\frac{20}{a_{2}}: \ (r_{41},r_{42},r_{43})=(-7,6,12),\nonumber\\
\phantom{\mbox{IV.b:\ \ }}y^{(4)}=a_{2}\left[y'y''+\frac{1}{5}a_{2}\left(y^{2}y''+yy'{}^{2}
-\frac{1}{125}a_{2}^{2}y^{5}\right)\right],
\label{eq:214}\end{gather}
\begin{gather}
\mbox{IV.c:\ \ }(p_{1},p_{2},p_{3},p_{4})=(36,36,-144,-144):\nonumber\\
\phantom{\mbox{IV.c:\ \ }}y_{01}^{2}=\frac{10}{a_{3}}:\ (r_{11},r_{12},r_{13})=(2,3,6),\nonumber\\
\phantom{\mbox{IV.c:\ \ }}y_{02}=-y_{01}: \
(r_{21},r_{22},r_{23})=(2,3,6),\nonumber
\end{gather}
\begin{gather}
 \phantom{\mbox{IV.c:\ \
}}y_{03}^{2}=\frac{40}{a_{3}}:\
(r_{31},r_{32},r_{33})=(-3,6,8), \nonumber\\
\phantom{\mbox{IV.c:\ \ }}y_{04}=-y_{03}: \ (r_{41},r_{42},r_{43})=(-3,6,8),\nonumber\\
\phantom{\mbox{IV.c:\ \ }}y^{(4)}=a_{3}\left(y^{2}y''+yy'{}^{2}-\frac{3}{50}a_{3}y^{5}\right),
\label{eq:215}\end{gather}
\begin{gather}
\mbox{IV.d:\ \ }(p_{1},p_{2},p_{3},p_{4})=(20,-120,-60,60):\nonumber\\
\phantom{\mbox{IV.d:\ \ }}y_{01}=\frac{2}{a_{1}}: \ (r_{11},r_{12},r_{13})=(1,2,10),\nonumber\\
\phantom{\mbox{IV.d:\ \ }}y_{02}=-\frac{8}{a_{1}}: \ (r_{21},r_{22},r_{23})=(-10,1,12),\nonumber\\
\phantom{\mbox{IV.d:\ \ }}y_{03}=\frac{4}{a_{1}}:\ (r_{31},r_{32},r_{33})=(-2,2,15),\nonumber\\
\phantom{\mbox{IV.d:\ \ }}y_{04}=-\frac{6}{a_{1}}:\ (r_{41},r_{42},r_{43})=(-3,-2,10),\nonumber\\
\phantom{\mbox{IV.d:\ \ }}y^{(4)}=a_{1}\!\left(yy'''\!-\frac{17}{2}y'y''\!+\frac{11}{4}a_{1}y^{2}y''\!
-\frac{15}{4}a_{1}yy'{}^{2}\!
+\frac{1}{2}a_{1}^{2}y^{3}y'\!
-\frac{1}{16}a_{1}^{3}y^{5}\right).\!\!
\label{eq:216}\end{gather}
The simplified equation for the case IV.d does not pass the
Painlev\'{e} test. So this case is
not be considered. The canonical
forms for the other cases can be obtained by adding the nondominant
terms with the coefficients $A_{k}(z)$,
$k=1,\ldots,12$ to the simplified equations.
All  coefficients $A_{k}$ can be obtained by
using the linear transformation (\ref{eq:36})
and the compatibility conditions at the resonances.
The canonical forms are as follows:

{\bf IV.a:}
\begin{gather}
y^{(4)}=-5yy'''-10\left(y'y''+y^{2}y''+y^{3}y'\right)-15y{y'}^{2}-y^{5}
+A_{1}\left(y'''+4yy''+3y'{}^{2}\right.\nonumber\\
\left.\phantom{y^{(4)}=}{}+6y^{2}y'+y^{4}\right)
+A_{3}\left(y''+3yy'+y^{3}\right)+A_{7}\left(y'+y^{2}\right)+A_{11}y+A_{12}.
\label{eq:217}\end{gather}
If one lets $y=u'/u$, (\ref{eq:217}) gives the
fifth order linear equation for $u$.
Equation (\ref{eq:217}) was  given in~\cite{cos2}.

{\bf IV.b:}
\begin{equation}
y^{(4)}=-5y'y''+5y^{2}y''+5yy'{}^{2}-y^{5}+(c_{1}z+c_{2})y+c_{3},
\label{eq:219}
\end{equation}
where $c_{i}$ are constants. The result (\ref{eq:219}) was
 given in \cite{cos2}.

{\bf IV.c:}
\begin{equation}
y^{(4)}=10y^{2}y''+10yy'{}^{2}-6y^{5}+c_{1}\left(y''-2y^{3}\right)+(c_{2}z+c_{3})y+c_{4},
\label{eq:220}
\end{equation}
where $c_{i}$ are constants. The result (\ref{eq:220}) was
 given in \cite{cos2, kudr}.

\section{Fifth order equations: $\boldsymbol{\mbox{P}_{\rm II}^{(5)}}$}
Differentiation of (\ref{eq:140}) and addition of the term $y^{6}$
which is also of order $-6$ as
$z \rightarrow z_{0}$ gives the following simplified equation of order five
\begin{gather}
y^{(5)}=a_{1}yy^{(4)}+a_{2}y'y'''+a_{3}y''{}^{2}
+a_{4}y^{2}y'''+a_{5}yy'y''+a_{6}y'{}^{3}\nonumber\\
\phantom{y^{(5)}=}{}+a_{7}y^{3}y''+
a_{8}y^{2}y'{}^{2}+a_{9}y^{4}y'+a_{10}y^{6},
\label{eq:230}\end{gather}
where $a_{i}$, $i=1,\ldots,10$, are constants. Substitution of
(\ref{eq:30}) into (\ref{eq:230})
into the equation above gives the following equations for
resonance $r$ and for
$y_0$ respectively,
\begin{gather}
Q(r)=(r+1)\Big\{r^{4}-(16+a_{1}y_{0})r^{3}-\left[a_{4}y_{0}^{2}
-(11a_{1}+a_{2})y_{0}-101\right]r^{2}\nonumber\\
\phantom{Q(r)=}{}-\left[a_{7}y_{0}^{3}-(7a_4 +a_5)y_0 +(46a_{1}+7a_{2}
+4a_{3})y_{0}+326\right]r\nonumber\\
\phantom{Q(r)=}{}-\left[a_{9}y_{0}^{4}-2(2a_{7}
+a_{8})y_{0}^{3}
+3(6a_{4}+2a_{5}+a_{6})y_{0}^{2}\right.\nonumber\\
\left.\phantom{Q(r)=}{}-8(12a_{1}+3a_{2}+2a_{3})y_{0}-600\vphantom{{}^2}\right]\Big\}=0,
\label{eq:231}\\
a_{10}y_{0}^{5}-a_{9}y_{0}^{4}+(2a_{7}+a_{8})y_{0}^{3}
-(6a_{4}+2a_{5}+a_{6})y_{0}^{2}\nonumber\\
\phantom{Q(r)=}{}+2(12a_{1} +3a_{2}+2a_{3})y_{0}+120=0.
\label{eq:231-1}
\end{gather}
Equation (\ref{eq:231-1}) implies that there are five branches if
$a_{6} \neq 0$. If $(r_{j1},r_{j2},r_{j3},r_{j4})$,
$j=1,\ldots,5,$ are the distinct integer resonances corresponding
to $(-1,y_{0j})$ and if $\prod\limits_{i=1}^{4}r_{ji}=p_{j}$,
where $p_{j}$ are integers and at least one of them is positive,
then the $p_{j}$ satisfy the following Diophantine equation,
\begin{equation}
\sum _{j=1}^{5} \frac{1}{p_{j}}=\frac{1}{120}.
\label{eq:232}
\end{equation}
The determination of all integer solutions $(p_{1},p_{2},p_{3},p_{4},p_{5})$
of the Diophantine equation is quite
difficult.
So, for the sake of completeness,
in this section we  present
special cases such as  single, double and triple branch cases.
We  also give an example
for the case of the four branches.
Since the procedure to obtain the canonical form
of the differential equations is the same as described
in the previous sections, we  only give the
canonical form of the differential equations for each cases.

The canonical form of the equation can be obtained by
adding the nondominant terms $y^{(4)}$, $yy'''$,
$y'y''$, $y^{2}y''$, $yy'{}^3$, $y^{3}y'$,
$y^{5}$, $y'''$, $yy''$, $y'{}^{2}$,
$y^{2}y'$, $y^{4}$, $y''$, $yy'$, $y^{3}$, $y'$, $y^{2}$, $y$, $1$ with
coefficients $A_{k}(z)$, $k=1,\ldots,19,$ which are analytic
functions of~$z$, respectively.

{\bf Case I.} If $a_{l}=0$, $l=4,\ldots,10$,  there is only
one branch and there are two cases such that the
resonances are distinct positive integers.

{\bf I.a:}
\begin{gather}
y_{01}=-2/a_{1}: \ (r_{11},r_{12},r_{13},r_{14})=(2,3,4,5),\nonumber\\
y^{(5)}=-2yy^{(4)}-8y'y'''-6y''{}^{2}+A_{1}\left(y^{4}+2yy''+6y'y''\right)
+A_{8}\left(y''+2yy''+2y'{}^{2}\right)\nonumber\\
\phantom{y^{(5)}=}{}+A_{13}(y''+2yy')+A_{16}\left(y'+y^{2}\right)+A_{19},
\label{eq:233}
\end{gather}
where $A_{1}$, $A_{8}$, $A_{13}$, $A_{16}$, $A_{19}$ are arbitrary
analytic functions of~$z$. Equation (\ref{eq:233}) can be
linearized by letting $u=y'+y^{2}$.

{\bf I.b:}
\[
y_{01}=-12/a_{2}: \ (r_{11},r_{12},r_{13},r_{14})=(1,4,5,6).
\]
In this case the linear transformation and the compatibility
conditions  give $A_i =0$, $i=1,\ldots,7$,
$A_{11}=A_{12}=A_{15}=0$ and
\begin{equation}
A''_{9}-\frac{1}{2}A_{9}^{2}=0.
\label{eq:235}
\end{equation}
Depending on the solution of (\ref{eq:235}) there
are following two subcases.

{\bf I.b.i:} $A_{9}=0$. The canonical form of the equation is
\begin{equation}
y^{(5)}=-12y'y'''-12y''{}^{2}+(c_{1}z+c_{2})y''
+2c_{1}y'+\frac{1}{6}(c_{1}z+c_{2})^{2},
\label{eq:236}
\end{equation}
where $c_{1}$ and $c_{2}$ are constants. If $c_{1} \neq 0,$
(\ref{eq:236}) can be reduced to (\ref{eq:42}). If
$c_{1}=0,$  (\ref{eq:236}) can be reduced to a third
order equation which belongs to the
hierarchy of the first Painlev\'{e} equation,
$\mbox{P}_{\rm I}^{(3)}$~\cite{mugan-fahd},
by integration once and letting $y=u'$.

{\bf I.b.ii:} $A_{9}=12/z^{2}$. The canonical
form of the equation is
\begin{gather}
y^{(5)}=-12y'y'''-12y''{}^{2}
+\frac{12}{z^2}\left(\frac{3}{2}y'''+yy''+2y'{}^{2}\right)
+\left(c_{1}z^{3}+\frac{c_{2}}{z^{2}}
-\frac{24}{z^{3}}\right)y'''\nonumber\\
\phantom{y^{(5)}=}{}-\frac{48}{z^{3}}yy'+\left(6c_{1}z^{2}
-\frac{4c_{2}}{z^{3}}+\frac{24}{z^{4}}\right)y'
+\left(4c_{1}+\frac{4c_{2}}{z^{4}}\right)y\nonumber\\
\phantom{y^{(5)}=}{}+\frac{24}{z^{4}}y^{2}
+\frac{1}{6}\left(c_{1}z^{3}+\frac{c_{2}}{z^{2}}\right)^{2},
\label{eq:237}
\end{gather}
where $c_{1}$ and $c_{2}$ are constants.

{\bf Case II.} $a_{7}=\cdots =a_{10}=0$: In this case
there are two branches. The resonances and the
canonical form of the equation are
\begin{gather}
y_{01}=\frac{-3}{a_1}:\ (r_{11},r_{12},r_{13},r_{14})=(1,3,4,5),\nonumber\\
y_{02}=\frac{-6}{a_1}: \ (r_{21},r_{22},r_{23},r_{24})=(-2,3,4,5),\nonumber\\
y^{(5)}=-3yy^{(4)}-12y'y'''-9y''{}^{2}-18yy'y''
-6y'{}^{3}-3y^{2}y''+(Ry)'''\nonumber\\
\phantom{y^{(5)}=}{}+
\frac{1}{3}A_{10}\left[y'''+3yy''+3y^{2}y'+3y'{}^{2}-(Ry)'\right]\nonumber\\
\phantom{y^{(5)}=}{}+A_{15}\left(y''+3yy'+y^{3}-Ry\right)+A_{19},
\label{eq:238}
\end{gather}
where $R=A_{8}-A_{9}/3$ and $A_{8}$, $A_{9}$, $A_{10}$,
$A_{15}$ and $A_{19}$
are arbitrary analytic functions of $z$. Equation (\ref{eq:238})
can be linearized if one lets
\begin{equation}
u=y''+3yy'+y^{3}-Ry.
\label{eq:239}
\end{equation}

{\bf Case III.} $a_{9}=a_{10}=0$: In this case there
are three branches. The resonances and the
canonical form of the equations are as follows:

{\bf III.a:}
\begin{gather}
y_{01}=\frac{-2}{a_{1}}:\
(r_{11},r_{12},r_{13},r_{14})=(2,3,4,5),\nonumber\\
y_{02}=-\frac{1-n}{a_{1}}: \
(r_{21},r_{22},r_{23},r_{24})=(4,5,6,n),\nonumber\\
y_{03}=-\frac{1+n}{a_{1}}: \
(r_{31},r_{32},r_{33},r_{34})=(4,5,6,-n),\nonumber\\
y^{(5)}=-2yy^{(4)}+\frac{1}{n^{2}-1}
\left[\left(56-8n^{2}\right)y'y'''+\left(54-6n^{2}\right)y''{}^{2}+48y^{2}y''\right.\nonumber\\
\phantom{y^{(5)}=}{}+\left.288yy'y''+96\left(y'{}^{3}+y^{3}y''\right)
+288y^{2}y'{}^{2}\right]+A_{8}\left(y'''+2yy''+y'{}^{2}\right) \nonumber\\
\phantom{y^{(5)}=}{}+(2A'_{8}+c_{1}z+c_{2})(y''+2yy')
+(A''_{8}+2c_{1})\left(y'+y^{2}\right)+A_{19},
\label{eq:240}
\end{gather}
where $A_{8}$ is an arbitrary analytic function of $z$ and
\begin{equation}
A_{19}=-\frac{n^{2}-1}{48}\left[A'''_{8}-A_{8}A_{8}''
-A'{}^{2}_{8}-A'_{8}(c_{1}z+c_{2})-2c_{1}A_{8}
+2(c_{1}z+c_{2})^{2}\right],
\label{eq:241}
\end{equation}
and $c_{1},c_{2}$ are constants,
$n \in \mathbb{Z}_{+}$, $n \neq 1,4,5,6$.
If $c_{1}=c_{2}=0$, double integration
of (\ref{eq:240}) yields~(\ref{eq:121}).

{\bf III.b:} The resonances are
\begin{gather}
y_{01}=-\frac{6-n}{a_{1}}: \ (r_{11},r_{12},r_{13},r_{14})=
(1,4,5,n),\nonumber\\
y_{02}=-\frac{6+n}{a_{1}}:\
(r_{21},r_{22},r_{23},r_{24})=(1,4,5,-n),\nonumber\\
y_{03}=-\frac{12}{a_{1}}: \
(r_{31},r_{32},r_{33},r_{34})=(-3,-2,4,5),
\label{eq:242}
\end{gather}
where $n \in \mathbb{Z}_{+} $ and $n \neq 1,4,5$.
It should be noted that, when $n \geq 6$, the double
integration of the simplified equation gives the
third order equation (\ref{eq:115}) with an additional
term $(c_{1}z+c_{2})$. Therefore the simplified
equation is not of Painlev\'{e} type
if $c_{1}$ and $c_{2} \neq 0$. Hence
we  only consider the cases for $n=2,3$.
The canonical form of the equation for $n=2$ is
\begin{gather}
y^{(5)}=2yy^{(4)}+7y'y'''+5y''{}^{2}
-\frac{3}{2}y^{2}y'''-9yy'y''
-3y'{}^{3}+\frac{1}{2}y^{3}y''\nonumber\\
\phantom{y^{(5)}=}{}+\frac{3}{2}y^{2}y'{}^{2}+A_{1}\left(y^{(4)}
-2yy'''-5y'y''+\frac{3}{2}y^{2}y'+3yy'{}^{2}-
\frac{1}{2}y^{3}y'\right)\nonumber\\
\phantom{y^{(5)}=}{}+A_{8}\left(y'''-2yy''-\frac{3}{2}y'{}^{2}
+\frac{3}{2}y^{2}y'-\frac{1}{8}y^{4}\right)
+A_{13}y''\nonumber\\
\phantom{y^{(5)}=}{}+(2A'_{13}-A_{1}A_{13})y'
+(A''_{13}-A_{1}A'_{13}-A_{8}A_{13})y+A_{19},
\label{eq:243}
\end{gather}
where $A_{1}$, $A_{8}$, $A_{13}$ and $A_{19}$ are arbitrary
analytic functions of~$z$. Double integration of~(\ref{eq:243})
yields  (\ref{eq:117}).

For $n=3$
\begin{gather}
y^{(5)}=2yy^{(4)}+\frac{1}{3}\left(26y'y'''+20y''{}^{2}
-\frac{16}{3}y^{2}y'''-32yy'y''
-\frac{32}{3}y'{}^{3}+\frac{16}{9}y^{3}y''\right.\nonumber\\
\left.\phantom{y^{(5)}=}{} +\frac{16}{3}y^{2}y'{}^{2}\right)
+A_{1}\left[y^{(4)}-2yy'''-\frac{2}{3}\left(10y'y''
-\frac{8}{3}y^{2}y''-\frac{16}{3}yy'{}^{2}+\frac{8}{9}y^{3}y'\right)\right]\nonumber\\
\phantom{y^{(5)}=}{}+A_{8}\left(y'''-2yy''-\frac{7}{3}y'{}^{2}
+\frac{16}{9}y^{2}y'-\frac{4}{27}y^{4}\right)+A_{19},
\label{eq:244}
\end{gather}
where $A_{1}$, $A_{8}$ and $A_{19}$ are arbitrary
analytic functions of $z$. If one lets
\begin{equation}
u=y'''-2yy''-\frac{7}{3}y'{}^{2}+\frac{16}{9}y^{2}y'-
\frac{4}{27}y^{4},
\label{eq:245}
\end{equation}
(\ref{eq:244}) can be reduced to a
linear equation for $u$.
It should be noted that (\ref{eq:245}) belongs
to $\mbox{P}_{\rm II}^{(3)}$ and  is given by  (\ref{eq:118}).

{\bf Case IV.} $a_{10}=0$: In this case there are
four branches and we will only give the following
 example.
\begin{gather}
y_{01}=-\frac{5}{a_{1}}: \
 (r_{11},r_{12},r_{13},r_{14})=(1,2,3,5),\nonumber\\
y_{02}=-\frac{10}{a_{1}}: \  (r_{21},r_{22},r_{23},r_{24})=(-2,1,2,5),\nonumber\\
y_{03}=-\frac{15}{a_{1}}: \
 (r_{31},r_{32},r_{33},r_{34})=(-3,-2,1,5),\nonumber\\
y_{04}=-\frac{20}{a_{1}}: \
(r_{41},r_{42},r_{43},r_{44})=(-2,-3,-4,5),\nonumber\\
y^{(5)}=-5\left(yy^{(4)}+3y'y'''+2y''{}^{2}+2y^{2}y'''+10yy'y''+3y'{}^{3}
+2y^{3}y''+6y^{2}y'{}^{2}+y^{4}y'\right)\!\nonumber\\
\phantom{y^{(5)}=}{}+A_{1}\left(y^{(4)}+5yy'''+10y'y''
+10y^2 y'+15yy'{}^{2}+10y^{3}y'+y^{5}\right)\nonumber\\
\phantom{y^{(5)}=}{}+A_{13}(y''+2yy')+A_{16}y'+(A'_{13}-A_{13}A_{1})y^{2}\nonumber\\
\phantom{y^{(5)}=}{}+(A'_{16}-A''_{13}+A'_{1}A_{13}
+2A_{1}A'_{13}-A_{1}A_{16}-A_{1}^{2}A_{13})+A_{19},
\label{eq:247}
\end{gather}
where $A_{1}$, $A_{13}$, $A_{16}$ and $A_{19}$ are arbitrary
analytic functions of $z$. Integration of~(\ref{eq:247}) once
gives the special case of (\ref{eq:217}).

In the procedure used to obtain higher order
 equations of Painlev\'{e} type,
the existence of at least one principal branch
has been imposed. But, the
compatibility conditions at the positive
resonances for the secondary branches
are identically satisfied for each cases.
Instead of having positive distinct
integer resonances, one can consider the case
of distinct integer resonances. In this case
it is possible to obtain equations like Chazy's equation~(\ref{eq:2})
which has three negative distinct integer resonances.
Starting from the first and second Painlev\'{e} equations and by using
this procedure one can look for polynomial type equations  of any order
of Painlev\'{e} type having  at least one principal branch.

\label{mugan-lastpage}
\end{document}